\documentclass[reqno]{amsart}
\usepackage{graphics,amsmath,color}
\usepackage[english]{babel}
\usepackage{latexsym}
\usepackage{amssymb}
\usepackage{graphpap}
\usepackage{indentfirst}
\usepackage[normalem]{ulem}
\usepackage{t1enc}            
\usepackage[utf8]{inputenc} 

\usepackage{geometry} 
\usepackage{pdflscape} 

\usepackage{bm}

\begin{document}

\title[\tiny{Generalized ballistic-conductive heat transport laws in three-dimensional isotropic materials}]{Generalized ballistic-conductive heat transport laws in three-dimensional isotropic materials}
\author[A. Fam\`a, L. Restuccia and P. V\'an]{A. Fam\`a$^{1}$, L. Restuccia$^{1}$ and P. V\'an$^{2,3,4}$}
\address{$^1$  University of Messina, Department of Mathematical and Computer Sciences, Physical Sciences and Earth Sciences 
$^2$Department of Theoretical Physics, Wigner Research Centre for Physics, H-1525 Budapest, Konkoly Thege Miklós u. 29-33., Hungary; //
and  $^3$Department of Energy Engineering, Faculty of Mechanical Engineering,  Budapest University of Technology and Economics, 1111 Budapest, Műegyetem rkp. 3., Hungary }
 
\date{\today}

\begin{abstract}
General constitutive equations of heat transport with second sound and ballistic propagation in isotropic materials are given using Non-Equilibrium Thermodynamics with Internal Variables (NET-IV). 
The consequences of Onsager reciprocity relations between thermodynamic fluxes and forces and positive definiteness of the entropy production are considered
. The relation to theories of Extended Thermodynamics is discussed in detail. We provide an explicit expression for all the components of the matrices of the transport coefficients. The expressions are cumbersome but are expected to be useful for computer programming for simulations of the corresponding physical effects. 
\end{abstract}
\maketitle

\section{Introduction}

There are several generalisations of classical Fourier law conduction that can also model second-sound phenomena (heat waves) and ballistic propagation. These theories are more and more important in nanostructures and are subjects of various challenging physical, mathematical and numerical researches. For example nonlocal effects and the role of effective temperature is investigated in \cite{Sob14a,Sob16a,Sob17a,Sob18a,SelEta17a}, particular special functions were constructed and exact solutions were calculated for both the hyperbolic and Guyer-Krumhansl heat conduction  \cite{Zhu16a1,Zhu17a,Zhu17a1,ZhuEta18a}, adapted numerical methods were developed in \cite{RieEta18a,NieCao19a}, the role of internal variables in complex media modelling were investigated in \cite{Res16a,CiaRes16a,CiaREs19a}, the particularities of heat conduction in nanomaterials is discovered in \cite{VazRio12a,CarEta19a,Mac19b}. These investigations are often related to various concepts of non-equilibrium temperature, too.

Second sound, the wavelike propagation of heat, is due to the inertia of internal energy. This property can be modelled by an additional non-equilibrium thermodynamic state variable. A straightforward choice for this additional vectorial state variable is the heat flux \cite{Mul67a1, Gya77a}. This choice leads to theories of Extended Thermodynamics (ET). There one requires a compatibility with  kinetic theory \cite{JouAta92b,LebEta08b,MulRug98b,CimEta14a,
Van16a,SelEta16b}, and the structure of the continuum theory will be compatible with the equations derived by moment series expansion of the Boltzmann equation, considering also a Callaway collision integral with two relaxation times. This compatibility with kinetic theory is a necessity for any phenomenology: a universal macroscopic approach must be valid in case of various micro- and mesostructures, in particular, it must be compatible with the theory of rarefied gases. 

The key of universality is to introduce only general physical and mathematical requirements and a minimal number of assumptions regarding the structure of the material. In particular, one must use and exploit the second law of thermodynamics and introduce a proper functional characterisation of the deviation from local equilibrium. All these can be accomplished most conveniently with the help of internal variables. 

One can achieve the compatibility with kinetic theory if the variables have the same tensorial order than the corresponding moments; therefore, their tensorial order is increasing with every new variable. However, the evolution equations of these fields are direct consequences of the second law, and one can get them solving the inequality of the entropy production. This way, for heat transport one obtains the Maxwell-Cattaneo-Vernotte equation as well as the Guyer-Krumhansl one with a single vectorial internal variable \cite{Van01a2, VanFul12a}. With an additional tensorial variable, a more general theory can be derived, that correctly describes ballistic propagation and the propagation of heat with the speed of sound, too \cite{KovVan15a}. 

Non-Equilibrium Thermodynamics with Internal Variables (NET-IV) can reproduce NaF experiments quantitatively, including the correct ballistic propagation speed \cite{KovVan16a,KovVan18a}. Nevertheless, the universality of the derivation indicates a broader range of validity, beyond rarefied real or phonon gases. This broadened range of validity is a prediction: e.g. one can expect non-Fourier heat transport in heterogeneous materials, too. Really, Guyer-Krumhansl type heat transport has been observed in diverse systems, in various heterogeneous materials with heat pulse experiments at room temperature \cite{BotEta16a, VanEta17a}. Internal variables are powerful for modelling concepts in other continuum theories, like rheology \cite{Ver97b, SzuFul18m}, semiconductor crystals with dislocations \cite{JouRes18a1}, porous nanocrystals filled by fluid flow \cite{Res10a, Res16a1, ResEta19a1, ResEta19a2}, and also in the GENERIC framework \cite{Ott05b}. Naturally, the relation of NET-IV with theories of ET, and kinetic theory, is not straightforward and its performance is analysed considering the complete theory, not only heat transport \cite{RugSug15b, RogEta18a, KovEta18m}. 

Up to now, the solutions and analyses of wave-like and ballistic propagation are mostly restricted to one spatial dimension. This approach is problematic from the point of view of experimental observations, especially considering the NaF experiments \cite{McNEta70a, JacWal71a}. In the classical experiments, the setup is not one-dimensional, but this fact is not considered in the usual modelling calculations \cite{KovVan16a, KovVan18a}. The related ET theory inherits the dimensional reduction from the particular collision integrals, e.g. the deviatoric and spherical contributions in the evolution equation of the heat flux have the same coefficient in the usual form of the Guyer-Krumhansl equation \cite{MulRug98b}, and this is preserved in nonlinear theories, too \cite{SelEta16b}.

In this paper we give the complete three-dimensional form of the equations of a theory of heat transport in isotropic materials, with a second order tensorial internal variable \textbf{Q}, including the possible Onsager reciprocity relations and second law requirements for the transport coefficients. The cases, where \textbf{Q} has odd parity and even parity, are developed separately. Since higher-order effects are taken into account, and, since we are considering the full three-dimensional problem, the explicit expressions we provide in the Appendix are cumbersome. However, they are expected to be useful in computer programming and simulations. 

The paper is organised as follows. In the second Section the theoretical framework is outlined and the basic balances and constitutive equations are given in a linear anisotropic form for the media under consideration. In Sections $3$ and $4$ the isotropic form of the equations are first treated in general. Then Onsager reciprocity relations are imposed as additional requirements, the entropy production is derived, the conditions of its positive definiteness are discussed and the generalized ballistic-conductive heat transport laws in three-dimensional isotropic materials are worked out. In  Section $5$ the general evolution equations for the heat flux, \textbf{q}, and for \textbf{Q} are derived. The same for \textbf{Q} with odd and even parities together with the one dimensional case is given in Sections $6$ and $7$. The general one dimensional form is more general than in \cite{KovVan15a}, while the obtained special cases of Jeffrey type, Maxwell-Cattaneo-Vernotte and Fourier heat equations are the same. Then, the conclusions are formulated. A detailed matrix form of the conductivity matrix is given in the Appendix, when \textbf{Q} has odd parity, and the differences with respect to the case where \textbf{Q} has even parity, are discussed, including the transformation of the sixth-order tensor to a form suitable for the calculation of the positive definiteness of the coefficients.

\section{Basic equations of heat transport coupled with a tensorial internal variable} 

We consider the balance equations of a rigid heat conductor, i.e. the balance of internal energy and the balance of entropy
\begin{equation}
\label{balance1}
\rho\dot{e} + q_{i,i}=0, 
\end{equation} 
\begin{equation}\label{balance2}
\rho\dot{s}+ J_{i,i} =\sigma^{(s)}.
\end{equation} 

\noindent Here $\rho$ is the density, $e$ the specific internal energy, $q_i$ the current density of the internal energy, the heat flux, $s$ the specific entropy, and $J_i$ denotes the entropy flux. The $\sigma^s$ entropy production rate plays a central and constructive role in the theory. $i,j,k$ are spatial indices related to Descartes coordinates, but they can also be considered as abstract spatial indices of vectors and tensors in the sense that they do not refer to particular coordinates \cite{Pen04b}; however, it is convenient in case of higher than second-order tensors. A comma in lower indices is for spatial derivation, and upper dot denotes the substantial time derivative (e.g. $\dot e = \partial_t e + v^i e_{,i}$, where $\partial_t$ is the partial time derivative). In case of rigid conductors at rest, the relative velocity of the continuum is zero; therefore, the substantial time derivative is equal to the partial time derivative. Regarding the general usage of abstract indices in classical nonrelativistic continuum theories see, e.g. in \cite{Van17a,VanEta19a}.  

We introduce an additional internal variable $Q_{ij}$ (a second-order tensor) which will incorporate higher-order effects in heat transport. Its physical meaning is not necessary a priori. However, in order that the reader may set some intuitive feeling of it, it is worth saying that $Q_{ij}$ may be interpreted as the flux of the heat flux (see Ref. \cite{JouAta92b,SelEta16b}) in solids, as the pressure tensor in fluids (see Ref. \cite{JouAta92b, KovVan18a}), or as the gradient of the heat flux, but here we leave open its meaning since it could also have a structural information about the particular material. We assume that $Q_{ij}$ contributes to the entropy and the entropy flux. The entropy flux must be zero if $q_i$ and $Q_{ij}$ are zero, that is in local thermodynamic equilibrium in the absence of heat flux. Therefore its most general form can be given as  
\begin{equation}
\label{eqn:3}
J_i=b_{ij}q_j+B_{ijk}Q_{jk},
\end{equation} 
where the $b_{ij}$ and $B_{ijk}$ constitutive functions are the Ny\'iri multipliers, that conveniently represent the deviation from the local equilibrium form of the entropy flux, like their quadratic form in the entropy density \cite{Nyi91a1}. 
This can be expressed also in an additive form, as the K vector of Müller, \cite{Mul67a}, if $K_i =  (b_{ij} - \delta_{ij}/T) q_j+B_{ijk}Q_{jk}$. 

Expanding the entropy function  
$s(e, q_i, Q_{ij})$  up to second-order approximation around a local equilibrium state, we obtain
\begin{equation}
\label{eqn:entropy}
s(e, q_i, Q_{ij})=s^{(eq)}(e)-\frac{1}{2\rho}m_{ij}q_iq_j-\frac{1}{2\rho}M_{ijkl}Q_{ij}Q_{kl}.
\end{equation} 
The coefficients $m_{ij}$ and $M_{ijkl}$ have the following symmetries 
\[
m_{ij}=m_{ji}, \quad M_{ijkl}=M_{klij}.
\]

Note that \eqref{eqn:3} and \eqref{eqn:entropy} are valid for anisotropic systems too. For isotropic systems $m_{ij}$ and $M_{ijkl}$ in \eqref{eqn:entropy} would reduce to a  scalar and the three scalar components conjugate to the three scalar invariants of tensor $Q_{ij}$, respectively. Thermodynamic stability requires that the inductivity tensors, $m_{ij}$, $M_{ijkl}$ (see in \cite{Gya77a, MacOns53a}), are  positive definite and we assume that they are constant. The entropy production $\sigma^{(s)}$, formed by combining \eqref{balance2}, \eqref{eqn:3} and \eqref{eqn:entropy}, is   
\begin{equation}                                                                                                                                                                                                                                                       
\label{eqn:EN}
\begin{split}
&\;\; \rho\dot{s}+J_{i,i}=\sigma^{(s)} \\
&= \rho\frac{d s^{(eq)}}{de}\dot{e}-\frac{1}{2}m_{ij}\dot{q_i}q_j-\frac{1}{2}m_{ij}q_i\dot{q_j}-\frac{1}{2}M_{ijkl}\dot{Q}_{ij}Q_{kl} \\
& \quad -\frac{1}{2}M_{ijkl}Q_{ij}\dot{Q}_{kl}+b_{ij,i}q_j+b_{ij}q_{j,i}+B_{ijk,i}Q_{jk}+B_{ijk}Q_{jk,i} \\
&= \left(b_{ij}-\frac{1}{T}\delta_{ij}\right)q_{j,i}+\left(b_{ji,j}-m_{ij}\dot{q_j}\right)q_i +\left(B_{kij,k}-M_{ijkl}\dot{Q}_{kl}\right)Q_{ij}+B_{ijk}Q_{jk,i}\geq 0. 
\end{split}
\end{equation}
Inequality \eqref{eqn:EN} expresses the second law of thermodynamics. Following the procedures of non-equilibrium thermodynamics we obtain the following \textit{general three-dimensional anisotropic linear relations} between the thermodynamic fluxes $b_{ij}-\frac{1}{T}\delta_{ij}, b_{ji,j}-m_{ij}\dot{q_j}, B_{ijk}, B_{kij,k}-M_{ijkl}\dot{Q}_{kl}$ and forces $q_i,q_{j,i},Q_{ij},Q_{jk,i} $    
\begin{align}
\label{eqn:general1}
b_{ji,j}-m_{ij}\dot{q_j}&=L^{(1)}_{ij}q_j+L^{(1,2)}_{ijk}q_{j,k}+L^{(1,3)}_{ijk}Q_{jk}+L^{(1,4)}_{ijkl}Q_{jk,l} \\
\label{eqn:general2}
b_{ij}-\frac{1}{T}\delta_{ij}&=L^{(2,1)}_{ijk}q_k+L^{(2)}_{ijkl}q_{k,l}+L^{(2,3)}_{ijkl}Q_{kl}+L^{(2,4)}_{ijklm}Q_{kl,m} \\
\label{eqn:general3}
B_{kij,k}-M_{ijkl}\dot{Q}_{kl}&=L^{(3,1)}_{ijk}q_k+L^{(3,2)}_{ijkl}q_{k,l}+L^{(3)}_{ijkl}Q_{kl}+L^{(3,4)}_{ijklm}Q_{kl,m} \\
\label{eqn:general4}
B_{ijk}&=L^{(4,1)}_{ijkl}q_l+L^{(4,2)}_{ijklm}q_{l,m}+L^{(4,3)}_{ijklm}Q_{lm}+L^{(4)}_{ijklmn}Q_{lm,n}. 
\end{align}

Here the conductivity tensors, $\textbf{L}^{(\alpha,\beta)}$ and $\textbf{L}^{(\gamma)}$, are restricted by material symmetries and by the second law. Furthermore reciprocity relations are also to be considered, as we do in Section $3$.

\section{Onsager reciprocity relations}

There are two different justifications of Onsager reciprocity. These are the assumptions regarding microscopic and macroscopic reversibility \cite{Mei75a}. The concept of microscopic reversibility goes back to Onsager, \cite{Ons31a1, Ons31a2}, and assumes a known microstructure, based on the reversal of microscopic velocities. The principle of macroscopic reversibility assumes a particular parity of the physical quantities regarding time reversal, which is originated in the consistency of the balances, and constitutive equations with a time reversal operation \cite{Ver97b, K1
}. Then the physical quantities with even parity are called $\alpha$- and with odd parity as $\beta$-type variables. For example density, entropy, energy and all thermostatic state variables are of $\alpha$-type, the velocity, heat flux, entropy flux are $\beta$-type, as one can see from the balances because time derivative changes the parity of the fields (e.g. the time derivative of an $\alpha$-type variable becomes $\beta$-type), but the gradient does not. It is generally assumed, that if the thermodynamic forces are of the same type, then the conductivity tensor is symmetric and when they are of the opposite, then the conductivity tensor becomes antisymmetric.

Several theoretical and experimental results support, that internal variable related thermodynamic fluxes and forces do not have definite parities, and both symmetric and antisymmetric parts of the conductivity tensors can be observed \cite{AssEta15a, VanEta14a, BerVan17b}. This is understandable because nothing is assumed about the microscopic structure of the material nor on the physical meaning of $Q_{ij}$ in NET-IV \cite{VanAta08a}. Therefore the microscopic reversibility conditions of Onsager cannot be applied, and  concept of macroscopic reversibility is not violated, if we assume that the internal variable, $Q_{ij}$, does not have parity. In the following, we start with the general case, without Onsagerian reciprocity and without any assumption on the parity of the $Q_{ij}$. Then we investigate the parities separately with symmetric and antisymmetric conductivity tensors. Let us remark, that comparison with Extended Thermodynamics identifies $Q_{ij}$  as a pressure tensor or as a flux of the heat flux \cite{KovEta18m}. In this case, it must have an even character, also because the entropy flux $J_i$ and the heat flux $q_i$ are $\beta$-type, odd quantities.

\subsection{Onsager reciprocity relations}
In this Subsection we suppose that the field $Q_{ij}$ (so also $Q_{ij,k}$) is \textit{odd} or \textit{even} functions under time reversal. 
Then $\left(B_{kij,k}-M_{ijkl}\dot{Q}_{kl}\right)$ and $B_{ijk}$ have an opposite parity, they are both \textit{even} or both \textit{odd} functions under time reversal. From this assumptions we obtain the following mathematical requirements (Onsager reciprocity relations) for the symmetric part of the conductivity tensor:
\begin{align}
\label{eqn:O-C1}
L^{(1)}_{ik}&= L^{(1)}_{ki}, & L^{(1,2)}_{ijk}&= L^{(2,1)}_{jki}, \\
\label{eqn:O-C2}
L^{(1,3)}_{ijk}&=\pm L^{(3,1)}_{jki}, & L^{(1,4)}_{ijkl}&=L^{(4,1)}_{jkli}, \\
\label{eqn:O-C3}
L^{(2)}_{ijkl}&= L^{(2)}_{klij}, & L^{(2,3)}_{ijkl}&=\pm L^{(3,2)}_{klij}, \\
\label{eqn:O-C4}
L^{(2,4)}_{ijklm}&=\pm L^{(4,2)}_{klmij}, & L^{(3)}_{ijkl}&= L^{(3)}_{klij}, \\
\label{eqn:O-C5}
L^{(3,4)}_{ijklm}&= L^{(4,3)}_{klmij}, & L^{(4)}_{ijklmn}&= L^{(4)}_{lmnijk}. 
\end{align} 
With the positive sign if $Q_{ij}$ is $\beta$-type and with a negative one if it is $\alpha$-type quantity. The sign is changes only if the parity of the respective thermodynamic forces changes, too. Therefore the the $\textbf{L}^{(\gamma)}$ tensors, that is the diagonal hypertensors in (\ref{eqn:general1})-(\ref{eqn:general4}) do not change sign.

\section{General isotropic case without assumption on the parity of \textbf{Q}}
In the general isotropic case, in which the symmetry properties of the body under consideration are invariant with respect to \textit{all rotations and to inversion of the frame of axes}, but in which Onsager reciprocity relations are not yet imposed, we have \cite{KeaFon75a}

\begin{gather}
\label{eqn:13}
m_{ij}=m\delta_{ij}, \\
\label{eqn:MM}
M_{ijkl}=M_1\delta_{ij}\delta_{kl}+M_2\delta_{ik}\delta_{jl}+M_3\delta_{il}\delta_{jk}, \\ 
\label{17}
L^{(1)}_{ij}\equiv \mathcal{L}^{(1)}_{ij}=L^{(1)}\delta_{ij}, \\
\label{eqn:1,4}
L^{(1,4)}_{ijkl}\equiv \mathcal{L}^{(1,4)}_{ijkl}=L^{(1,4)}_1\delta_{ij}\delta_{kl}+L^{(1,4)}_2\delta_{ik}\delta_{jl}+L^{(1,4)}_3\delta_{il}\delta_{jk}, \\
L^{(2)}_{ijkl}\equiv \mathcal{L}^{(2)}_{ijkl}=L^{(2)}_1\delta_{ij}\delta_{kl}+L^{(2)}_2\delta_{ik}\delta_{jl}+L^{(2)}_3\delta_{il}\delta_{jk}, \\
\label{eqn:2,3}
L^{(2,3)}_{ijkl}\equiv \mathcal{L}^{(2,3)}_{ijkl}=L^{(2,3)}_1\delta_{ij}\delta_{kl}+L^{(2,3)}_2\delta_{ik}\delta_{jl}+L^{(2,3)}_3\delta_{il}\delta_{jk}, \\
\label{eqn:3,2}
L^{(3,2)}_{ijkl}\equiv \mathcal{L}^{(3,2)}_{ijkl}=L^{(3,2)}_1\delta_{ij}\delta_{kl}+L^{(3,2)}_2\delta_{ik}\delta_{jl}+L^{(3,2)}_3\delta_{il}\delta_{jk}, \\
L^{(3)}_{ijkl}\equiv \mathcal{L}^{(3)}_{ijkl}=L^{(3)}_1\delta_{ij}\delta_{kl}+L^{(3)}_2\delta_{ik}\delta_{jl}+L^{(3)}_3\delta_{il}\delta_{jk}, \\
\label{eqn:4,1}
L^{(4,1)}_{ijkl}\equiv \mathcal{L}^{(4,1)}_{ijkl}=L^{(4,1)}_1\delta_{ij}\delta_{kl}+L^{(4,1)}_2\delta_{ik}\delta_{jl}+L^{(4,1)}_3\delta_{il}\delta_{jk}, \\
\begin{split}
\label{eqn:4,4}
L^{(4)}_{ijklmn}\equiv \mathcal{L}^{(4)}_{ijklmn}&=L^{(4)}_1\delta_{ij}\delta_{kl}\delta_{mn}+L^{(4)}_2\delta_{ij}\delta_{km}\delta_{ln}+L^{(4)}_3\delta_{ij}\delta_{kn}\delta_{lm} \\
& \quad +L^{(4)}_4\delta_{ik}\delta_{jl}\delta_{mn}+L^{(4)}_5\delta_{ik}\delta_{jm}\delta_{ln}+L^{(4)}_6\delta_{ik}\delta_{jn}\delta_{lm} \\
& \quad +L^{(4)}_7\delta_{il}\delta_{jk}\delta_{mn}+L^{(4)}_8\delta_{im}\delta_{jk}\delta_{ln}+L^{(4)}_9\delta_{in}\delta_{jk}\delta_{lm} \\
& \quad +L^{(4)}_{10}\delta_{il}\delta_{jm}\delta_{kn}+L^{(4)}_{11}\delta_{im}\delta_{jl}\delta_{kn}+L^{(4)}_{12}\delta_{in}\delta_{jl}\delta_{km} \\ 
& \quad +L^{(4)}_{13}\delta_{in}\delta_{jm}\delta_{kl}++L^{(4)}_{14}\delta_{im}\delta_{jn}\delta_{kl}+L^{(4)}_{15}\delta_{il}\delta_{jn}\delta_{km}.
\end{split}
\end{gather}
The coefficients appearing in the entropy \eqref{eqn:entropy} are \eqref{eqn:13} and \eqref{eqn:MM}. 

Furthermore, in the isotropic case (where the symmetry properties of the considered body are invariant only with respect to all rotations of the frame of axes) the third and fifth order tensors keep the form $L_{ijk}=L\in_{ijk}$ and $L_{ijklm}=A_1\in_{ijk}\delta_{lm}+A_2\in_{ijl}\delta_{km}+A_3\in_{ijm}\delta_{kl}+A_4\in_{ikl}\delta_{jm}+A_5\in_{ikm}\delta_{lj}+ \\ A_6\in_{ilm}\delta_{jk}$ respectively, where $\in_{ijk}$ denotes the Levi Civita tensor and the quantities $L$ and $A_i$, $i=1,\dots ,6$, are the independent components of the tensors $L_{ijk}$ and $L_{ijklmn}$, that vanish when there is also the invariance of the properties with respect to the inversion of the axes. Thus, we obtain
\begin{gather}
\label{eqn:30}
L^{(1,2)}_{ijk}=L^{(1,3)}_{ijk}=L^{(2,1)}_{ijk}=L^{(3,1)}_{ijk}=0, \\
\label{eqn:31}
L^{(2,4)}_{ijklm}=L^{(3,4)}_{ijklm}=L^{(4,2)}_{ijklm}=L^{(4,3)}_{ijklm}=0.
\end{gather}

From relations \eqref{eqn:13}-\eqref{eqn:31}, the phenomenological equations \eqref{eqn:general1}-\eqref{eqn:general4} in the isotropic case read
\begin{align}
\label{eqn:I1}
m\dot{q_i}-b_{ji,j}&=-L^{(1)}q_i-L^{(1,4)}_1Q_{ik,k}-L^{(1,4)}_2Q_{ki,k}-L^{(1,4)}_3Q_{kk,i}, \\
\begin{split}
\label{eqn:I2}
b_{ij}-\frac{1}{T}\delta_{ij}&=L^{(2)}_1\delta_{ij}q_{k,k}+L^{(2)}_2q_{i,j}+L^{(2)}_3q_{j,i}+L^{(2,3)}_1\delta_{ij}Q_{kk} \\
& \quad +L^{(2,3)}_2Q_{ij}+L^{(2,3)}_3Q_{ji}, 
\end{split} 
\\
\begin{split}
\label{eqn:I3}
B_{kij,k}&=M_1\delta_{ij}\dot{Q}_{kk}+M_2\dot{Q}_{ij}+M_3\dot{Q}_{ji}+L^{(3,2)}_1\delta_{ij}q_{k,k}+L^{(3,2)}_2q_{i,j} \\
& \quad +L^{(3,2)}_3q_{j,i}+L^{(3)}_1\delta_{ij}Q_{kk}+L^{(3)}_2Q_{ij}+L^{(3)}_3Q_{ji}, 
\end{split}
\\
\begin{split}
\label{eqn:I4}
B_{ijk}&=L^{(4,1)}_1\delta_{ij}q_k+L^{(4,1)}_2\delta_{ik}q_j+L^{(4,1)}_3\delta_{jk}q_i \\
& \quad +\delta_{ij}\left(L^{(4)}_1Q_{kl,l}+L^{(4)}_2Q_{lk,l}+L^{(4)}_3Q_{ll,k}\right) \\
& \quad +\delta_{ik}\left(L^{(4)}_4Q_{jl,l}+L^{(4)}_5Q_{lj,l}+L^{(4)}_6Q_{ll,j}\right) \\
& \quad +\delta_{jk}\left(L^{(4)}_7Q_{il,l}+L^{(4)}_8Q_{li,l}+L^{(4)}_9Q_{ll,i}\right) \\
& \quad +L^{(4)}_{10}Q_{ij,k}+L^{(4)}_{11}Q_{ji,k}+L^{(4)}_{12}Q_{jk,i}+L^{(4)}_{13}Q_{kj,i} \\
& \quad +L^{(4)}_{14}Q_{ki,j}+L^{(4)}_{15}Q_{ik,j}. 
\end{split}
\end{align}

In the general isotropy case the number of material coefficients of equations \eqref{eqn:I1}-\eqref{eqn:I4} are $38$: $4$ \textit{static} ($m$ and $M_i$) and $34$ independent \textit{conductivity parameters} ($\textbf{L}^{(\varepsilon, \mu)}$ and $\textbf{L}^{(\delta)}$).

\subsection{Onsager symmetry }
Now, we tentatively require Onsager reciprocity relations \eqref{eqn:O-C1}-\eqref{eqn:O-C5}, as additional restrictions on the coefficients, and explore which further reduction this implies on the number of independent conductivity parameters 
Then, from \eqref{eqn:O-C2}$_2$   $\mathcal{L}^{(1,4)}_{ijkl}=\mathcal{L}^{(4,1)}_{jkli}$,  and being 
\begin{equation}
\label{eqn:81}
\mathcal{L}^{(4,1)}_{jkli}=L^{(4,1)}_1\delta_{jk}\delta_{li}+L^{(4,1)}_2\delta_{jl}\delta_{ki}+L^{(4,1)}_3\delta_{ji}\delta_{kl},
\end{equation}
we  obtain 
\begin{equation}
\label{eqn:32}
L^{(1,4)}_1=\pm L^{(4,1)}_3, \quad L^{(1,4)}_2=\pm L^{(4,1)}_2, \quad L^{(1,4)}_3=\pm L^{(4,1)}_1.
\end{equation}

\medskip

Furthermore, for each  isotropic four tensor $\mathcal{L}_{ijkl}$ we have  the following symmetry relation 
\begin{equation}
\label{eqn:property}
\mathcal{L}_{ijkl}=\mathcal{L}_{klij},
\end{equation}

\noindent because of 
\begin{equation}
\mathcal{L}_{ijkl}=T_1\delta_{ij}\delta_{kl}+T_2\delta_{ik}\delta_{jl}+T_1\delta_{il}\delta_{jk}=T_1\delta_{kl}\delta_{ij}+T_2\delta_{ki}\delta_{lj}+T_3\delta_{kj}\delta_{li}=\mathcal{L}_{klij},
\end{equation} 

\noindent where $T_1$, $T_2$ and $T_3$ indicate the independent components of $\mathcal{L}_{ijkl}$. Taking into account the property \eqref{eqn:property}, Onsager relations $\eqref{eqn:O-C3}_1$ and $\eqref{eqn:O-C4}_2$ are  verified in the isotropic case and  from $\eqref{eqn:O-C3}_2$ we derive $\mathcal{L}^{(2,3)}_{ijkl}=\mathcal{L}^{(3,2)}_{klij}=\mathcal{L}^{(3,2)}_{ijkl}$, from which we have 
\begin{equation}
\label{eqn:33}
L^{(2,3)}_i=\pm L^{(3,2)}_i \quad  (i=1,2,3). 
\end{equation} 

\medskip

Then, from \eqref{eqn:4,4} we obtain
\begin{equation}
\begin{split}
\label{eqn:4,4-2}
\mathcal{L}^{(4)}_{lmnijk}&=L^{(4)}_1\delta_{lm}\delta_{ni}\delta_{jk}+L^{(4)}_2\delta_{lm}\delta_{nj}\delta_{ik}+L^{(4)}_3\delta_{lm}\delta_{nk}\delta_{ij}+L^{(4)}_4\delta_{ln}\delta_{mi}\delta_{jk} \\
& \quad +L^{(4)}_5\delta_{ln}\delta_{mj}\delta_{ik}+L^{(4)}_6\delta_{ln}\delta_{mk}\delta_{ij}+L^{(4)}_7\delta_{mn}\delta_{li}\delta_{jk}+L^{(4)}_8\delta_{mn}\delta_{lj}\delta_{ik} \\
& \quad +L^{(4)}_9\delta_{mn}\delta_{lk}\delta_{ij}+L^{(4)}_{10}\delta_{li}\delta_{mj}\delta_{nk}+L^{(4)}_{11}\delta_{lj}\delta_{mi}\delta_{nk}+L^{(4)}_{12}\delta_{lk}\delta_{mi}\delta_{nj} \\
& \quad +L^{(4)}_{13}\delta_{lk}\delta_{mj}\delta_{ni}+L^{(4)}_{14}\delta_{lj}\delta_{mk}\delta_{in}+L^{(4)}_{15}\delta_{li}\delta_{mk}\delta_{nj}.
\end{split}
\end{equation}
Adding \eqref{eqn:4,4} and \eqref{eqn:4,4-2}, using Onsager relation $\eqref{eqn:O-C5}_2$ and dividing by $2$, we have
\begin{equation}
\label{eqn:34}
\begin{split}
\mathcal{L}^{(4)}_{ijklmn}&=C^{(4)}_1(\delta_{ij}\delta_{kl}\delta_{mn}+\delta_{in}\delta_{jk}\delta_{lm})+C^{(4)}_2(\delta_{ij}\delta_{km}\delta_{ln}+\delta_{ik}\delta_{jn}\delta_{lm}) \\
& \quad +C^{(4)}_3\delta_{ij}\delta_{kn}\delta_{lm}+C^{(4)}_4(\delta_{ik}\delta_{jl}\delta_{mn}+\delta_{im}\delta_{jk}\delta_{nl})+C^{(4)}_5\delta_{ik}\delta_{jm}\delta_{ln} \\
& \quad +C^{(4)}_6\delta_{il}\delta_{jk}\delta_{mn}+C^{(4)}_7\delta_{il}\delta_{jm}\delta_{kn}+C^{(4)}_8\delta_{il}\delta_{jn}\delta_{km}+C^{(4)}_9\delta_{im}\delta_{jl}\delta_{kn} \\
& \quad +C^{(4)}_{10}(\delta_{im}\delta_{jn}\delta_{kl}+\delta_{in}\delta_{jl}\delta_{km})+C^{(4)}_{11}\delta_{in}\delta_{jm}\delta_{kl},
\end{split}
\end{equation}
where
\begin{gather}
\label{eqn:C1-3}
C^{(4)}_{1}=\frac{L^{(4)}_1+L^{(4)}_9}{2}, \quad C^{(4)}_{2}=\frac{L^{(4)}_2+L^{(4)}_6}{2}, \quad C^{(4)}_{3}=L^{(4)}_3, \\
C^{(4)}_{4}=\frac{L^{(4)}_4+L^{(4)}_8}{2}, \quad C^{(4)}_{5}=L^{(4)}_5, \quad C^{(4)}_{6}=L^{(4)}_7, \quad 
C^{(4)}_{7}=L^{(4)}_{10}, \\
\label{eqn:C8-11}
 C^{(4)}_{8}=L^{(4)}_{15}, \quad C^{(4)}_{9}=L^{(4)}_{11}, \quad C^{(4)}_{10}=\frac{L^{(4)}_{12}+L^{(4)}_{14}}{2}, \quad C^{(4)}_{11}=L^{(4)}_{13}. 
\end{gather}

Thus, from  relation $\mathcal{L}^{(4)}_{ijklmn}=\mathcal{L}^{(4)}_{lmnijk}$  the significant components of the isotropic tensor $\mathcal{L}_{ijklmn}$  reduce from $15$ to $11$. Therefore, in case that Onsager reciprocity is imposed, from relations \eqref{eqn:32}, \eqref{eqn:33} and \eqref{eqn:34} the number of conductivity parameters are reduced altogether from $34$ to $24$. 

\subsection{Entropy production} 
\label{4.2}
In the general isotropic case, with the aid of relations \eqref{eqn:I1}-\eqref{eqn:I4}, \eqref{eqn:13}-\eqref{eqn:4,1}, \eqref{eqn:30} and \eqref{eqn:31}, entropy production \eqref{eqn:EN} can be written as
\begin{equation}
\label{sigmas1}
\begin{split}
\sigma^{(s)}&=\mathcal{L}^{(1)}_{ik}q_iq_k+\mathcal{L}^{(2)}_{ijkl}q_{j,i}q_{k,l}+\mathcal{L}^{(3)}_{ijkl}Q_{ij}Q_{kl}+\mathcal{L}^{(4)}_{ijklmn}Q_{jk,i}Q_{lm,n}+\\
& \quad +\left(\mathcal{L}^{(1,4)}_{ijkl}+\mathcal{L}^{(4,1)}_{ljki}\right)q_iQ_{jk,l}+\left(\mathcal{L}^{(2,3)}_{ijkl}+\mathcal{L}^{(3,2)}_{klji}\right)q_{j,i}Q_{kl}\geq 0.
\end{split}
\end{equation}
In the case where the internal variable $Q_{ij}$ \textit{has odd parity}, using Onsager relations and \eqref{eqn:33} and \eqref{eqn:34}, expression \eqref{sigmas1} takes the form
\begin{equation}
\label{eqn:36}
\begin{split}
\sigma^{(s)}&=\mathcal{L}^{(1)}_{ik}q_iq_k+\mathcal{L}^{(2)}_{ijkl}q_{j,i}q_{k,l}+ \mathcal{L}^{(3)}_{ijkl}Q_{ij}Q_{kl}+ \mathcal{L}^{(4)}_{ijklmn}Q_{jk,i}Q_{lm,n} \\
& \quad +\left(\mathcal{L}^{(1,4)}_{ijkl}\pm \mathcal{L}^{(1,4)}_{iljk}\right)q_iQ_{jk,l} +\left(\mathcal{L}^{(2,3)}_{ijkl}\pm \mathcal{L}^{(2,3)}_{klji}\right)q_{j,i}Q_{kl} \geq 0, 
\end{split}
\end{equation}
or in extended form
\begin{equation}
\label{eqn:EI}
\begin{split}
\sigma^{(s)}&=L^{(1)}\delta_{ik}q_iq_k+\left(L^{(2)}_1\delta_{ji}\delta_{kl}+L^{(2)}_2\delta_{jk}\delta_{il}+L^{(2)}_3\delta_{jl}\delta_{ik}\right)q_{i,j}q_{k,l} \\
& \quad +\left(L^{(3)}_1\delta_{ij}\delta_{kl}+L^{(3)}_2\delta_{ik}\delta_{jl}+L^{(3)}_3\delta_{il}\delta_{jk}\right)Q_{ij}Q_{kl} \\
& \quad +\left[C^{(4)}_1(\delta_{pi}\delta_{jl}\delta_{mn}+\delta_{pn}\delta_{ij}\delta_{lm})+C^{(4)}_2(\delta_{pi}\delta_{jm}\delta_{ln}+\delta_{pj}\delta_{in}\delta_{lm})\right. \\
& \quad +C^{(4)}_3\delta_{pi}\delta_{jn}\delta_{lm}+C^{(4)}_4(\delta_{pj}\delta_{il}\delta_{mn}+\delta_{pm}\delta_{ij}\delta_{nl})+C^{(4)}_5\delta_{pj}\delta_{im}\delta_{ln} \\
& \quad +C^{(4)}_6\delta_{pl}\delta_{ij}\delta_{mn}+C^{(4)}_7\delta_{pl}\delta_{im}\delta_{jn}+C^{(4)}_8\delta_{pl}\delta_{in}\delta_{jm}+C^{(4)}_9\delta_{pm}\delta_{il}\delta_{jn} \\
& \quad +\left.C^{(4)}_{10}(\delta_{pm}\delta_{in}\delta_{jl}+\delta_{pn}\delta_{il}\delta_{jm})+C^{(4)}_{11}\delta_{pn}\delta_{im}\delta_{jl}\right]Q_{ij,p}Q_{lm,n} \\
& \quad +\left(L^{(1,4)}_1\delta_{il}\delta_{mn}+L^{(1,4)}_2\delta_{im}\delta_{ln}+L^{(1,4)}_3\delta_{in}\delta_{lm}\right)q_iQ_{lm,n} \\
& \quad \pm \left(L^{(1,4)}_1\delta_{kp}\delta_{ij}+L^{(1,4)}_2\delta_{ki}\delta_{pj}+L^{(1,4)}_3\delta_{kj}\delta_{pi}\right)Q_{ij,p}q_k \\
& \quad +\left(L^{(2,3)}_1\delta_{ji}\delta_{kl}+L^{(2,3)}_2\delta_{jk}\delta_{il}+L^{(2,3)}_3\delta_{jl}\delta_{ik}\right)q_{i,j}Q_{kl} \\
& \quad \pm \left(L^{(2,3)}_1\delta_{ij}\delta_{kl}+L^{(2,3)}_2\delta_{ik}\delta_{jl}+L^{(2,3)}_3\delta_{il}\delta_{jk}\right)Q_{ij}q_{k,l} \geq 0.
\end{split}
\end{equation}

From \eqref{eqn:EI} it is seen that the entropy production is a non-negative bilinear form in the components of the heat flux and its gradient, and in the components of the internal variable and its gradient (see in Appendix its matrix representation $\sigma^{(s)}=X_\alpha \mathcal{L}_{\alpha\beta}X_{\beta}$, with $X_\alpha$, $X_\beta$ and $\mathcal{L}_{\alpha\beta}$ suitable matrices).

The following inequalities can be obtained for the components of the phenomenological tensors, resulting from the fact that all the elements of the main diagonal of the symbolic matrix $\{\mathcal{L}_{\alpha\beta}\}$ associated to the bilinear form \eqref{eqn:EI} must be non-negative, representing a condition (only necessary) for the semi-definiteness of the matrix $\{\mathcal{L}_{\alpha\beta}\}$ (see Appendix)
\begin{gather}
\label{eqn:D1}
L^{(1)}\geq 0, \qquad L^{(2)}_3\geq 0, \qquad L^{(3)}_2\geq 0, \\[0.5em]
\label{eqn:D2}
L^{(2)}_1+L^{(2)}_2+L^{(2)}_3\geq 0, \qquad L^{(3)}_1+L^{(3)}_2+L^{(3)}_3\geq 0, \\[0.5em]
\begin{split}
\label{eqn:D3}
2C^{(4)}_1+2C^{(4)}_2+C^{(4)}_3+2C^{(4)}_4&+C^{(4)}_5+C^{(4)}_6+C^{(4)}_7+C^{(4)}_8 \\ 
 & +C^{(4)}_9+2C^{(4)}_{10}+C^{(4)}_{11}\geq 0,
\end{split} 
\\[0.5em]
\label{eqn:D4}
C^{(4)}_2+C^{(4)}_8+C^{(4)}_{10}\geq 0, \qquad C^{(4)}_4+C^{(4)}_9+C^{(4)}_{10}\geq 0, \\[0.5em]
\label{eqn:D5}
C^{(4)}_{10}\geq 0, \qquad C^{(4)}_1+C^{(4)}_{10}+C^{(4)}_{11}\geq 0.
\end{gather} 
Relations \eqref{eqn:D3}-\eqref{eqn:D5}, come from  the non-negativity of the elements of the main diagonal of the  sub-matrix $\mathcal{L}^{(4)}_{pijlmn}$. 

Moreover, other relations can be obtained from the non-negativity of the major minors $P_r$ ($r=1,\ldots ,48$) of $\{\mathcal{L}_{\alpha\beta}\}$, coming from Sylvester's criterion, that represents a necessary and sufficient condition for the semi-definiteness of the matrix $\{\mathcal{L}_{\alpha\beta}\}$. For instance, the calculation of the major minors up to sixth-order gives the relations $\eqref{eqn:D1}_1$, $\eqref{eqn:D1}_2$ and $\eqref{eqn:D2}_1$. 

\noindent The non-negativity of the seventh-order major minor of $\{\mathcal{L}_{\alpha\beta}\}$
\begin{equation}
P_7=
\begin{vmatrix}
L^{(1)} & 0 & 0 & 0 & 0 & 0 & 0 \\
0 & L^{(1)} & 0 & 0 & 0 & 0 & 0 \\
0 & 0 & L^{(1)} & 0 & 0 & 0 & 0 \\
0 & 0 & 0 & L^{(2)} & 0 & 0 & 0 \\
0 & 0 & 0 & 0 & L^{(2)}_3 & 0 & L^{(2)}_2 \\
0 & 0 & 0 & 0 & 0 & L^{(2)}_3 & 0 \\
0 & 0 & 0 & 0 & L^{(2)}_2 & 0 & L^{(2)}_3 \\
\end{vmatrix}
,
\end{equation}
with $L^{(2)}\equiv L^{(2)}_1+L^{(2)}_2+L^{(2)}_3$, gives the new relation
\begin{equation}
\label{50}
L^{(2)}_2+\left(L^{(2)}_3\right)^2\geq 0,
\end{equation}
and so on. In the Appendix we give a two-dimensional form of the conductivity matrix $\{\mathcal{L}_{\alpha\beta}\}$, in terms of which the calculation of the conditions of positive definiteness is straightforward.

\section{Rate equations for $q_i$ and $Q_{ij}$ in the general case without assumption on the parity of $Q_{ij}$}
\label{4.3}
Changing indexes $i$ and $j$ in \eqref{eqn:I2}, deriving it with respect to $x_j$ and substituting it into \eqref{eqn:I1}, we deduce
\begin{equation}
\label{eqn:A0}
\begin{split}
m\dot{q_i}+L^{(1)}q_i&=\left(L^{(2)}_1+L^{(2)}_2\right)q_{k,ki}+L^{(2)}_3q_{i,kk}+\left(L^{(2,3)}_1-L^{(1,4)}_3\right)Q_{kk,i} \\
& \quad +\left(L^{(2,3)}_3-L^{(1,4)}_1\right)Q_{ik,k}+\left(L^{(2,3)}_2-L^{(1,4)}_2\right)Q_{ki,k}+{\left(\frac{1}{T}\right)}_{,i}.
\end{split}
\end{equation}
where
\begin{gather}
m>0, \quad L^{(1)}>0, \quad L^{(2)}_1+L^{(2)}_2>0, \quad L^{(2)}_3>0.
\end{gather}
Equation \eqref{eqn:A0} can be written as follows

\begin{equation}
\label{eqn:A}
\tau\dot{q_i}+q_i=-\lambda T_{,i}+l_1q_{i,kk}+l_2q_{k,ki}+l_{12}Q_{kk,i}+l_{13}Q_{ik,k}+l_{14}Q_{ki,k}\ ,
\end{equation}
where 
\begin{gather}
\label{eqn:54}
\tau=\frac{m}{L^{(1)}}, \quad \lambda=\frac{1}{L^{(1)}T^2}, \quad l_1=\frac{L^{(2)}_3}{L^{(1)}}, \quad l_2=\frac{L^{(2)}_1+L^{(2)}_2}{L^{(1)}},\\
\label{eqn:55}
l_{12}=\frac{L^{(2,3)}_1-L^{(1,4)}_3}{L^{(1)}}, \quad l_{13}=\frac{L^{(2,3)}_3-L^{(1,4)}_1}{L^{(1)}}, \quad l_{14}=\frac{L^{(2,3)}_2-L^{(1,4)}_2}{L^{(1)}},
\end{gather}
being $\tau$ the relaxation time of the heat flux (that, then, has a finite velocity of propagation), $\lambda$ the heat conductivity and $l_i$ have dimension of square length. 

In analogous way, if we change $i\rightarrow k$, $j\rightarrow i$, $k\rightarrow j$ in equation \eqref{eqn:I4},  deriving it with respect to $x_k$ and  inserting it into \eqref{eqn:I3},  we have
\begin{equation}
\label{eqn:b}
\begin{split}
&M_1\delta_{ij}\dot{Q}_{kk}+M_2\dot{Q}_{ij}+M_3\dot{Q}_{ji}+L^{(3)}_1\delta_{ij}Q_{kk}+L^{(3)}_2Q_{ij}+L^{(3)}_3Q_{ji}\\
& \quad =\left(L^{(4,1)}_3-L^{(3,2)}_1\right)\delta_{ij}q_{k,k}+\left(L^{(4,1)}_2-L^{(3,2)}_2\right)q_{i,j}+\left(L^{(4,1)}_1-L^{(3,2)}_3\right)q_{j,i}\\
& \quad +\left(L^{(4)}_3+L^{(4)}_6\right)Q_{kk,ij}+L^{(4)}_{12}Q_{ij,kk}+L^{(4)}_{13}Q_{ji,kk}+\left(L^{(4)}_1+L^{(4)}_{15}\right)Q_{jk,ik}\\
& \quad +\left(L^{(4)}_2+L^{(4)}_{11}\right)Q_{kj,ik}+\left(L^{(4)}_4+L^{(4)}_{14}\right)Q_{ik,jk}+\left(L^{(4)}_5+L^{(4)}_{10}\right)Q_{ki,jk}\\
& \quad +\delta_{ij}\left[\left(L^{(4)}_7+L^{(4)}_8\right)Q_{kl,lk}+L^{(4)}_9Q_{ll,kk}\right],
\end{split}
\end{equation}
i.e.
\begin{equation}
\label{eqn:B}
\begin{split}
&\tau_1\delta_{ij}\dot{Q}_{kk}+\tau_2\dot{Q}_{ij}+\tau_3\dot{Q}_{ji}+\delta_{ij}Q_{kk}+l^3_2Q_{ij}+l^3_3Q_{ji}=l_{21}\delta_{ij}q_{k,k}+l_{31}q_{i,j}\\
& \quad +l_{41}q_{j,i}+L_1Q_{kk,ij}+L_2Q_{ij,kk}+L_3Q_{ji,kk}+L_4Q_{jk,ik}+L_5Q_{kj,ik} \\
& \quad +L_6Q_{ik,jk}+L_7Q_{ki,jk}+\delta_{ij}\left({L_8Q_{kl,kl}+L_9Q_{ll,kk}}\right),
\end{split}
\end{equation}
where 
\begin{gather}
\label{eqn:58}
\tau_1=\frac{M_1}{L^{(3)}_1}, \quad \tau_2=\frac{M_2}{L^{(3)}_1}, \quad \tau_3=\frac{M_3}{L^{(3)}_1}, \quad l^3_2=\frac{L^{(3)}_2}{L^{(3)}_1}, \quad l^3_3=\frac{L^{(3)}_3}{L^{(3)}_1}, \\
\label{eqn:59}
l_{21}=\frac{L^{(4,1)}_3-L^{(3,2)}_1}{L^{(3)}_1}, \quad l_{31}=\frac{L^{(4,1)}_2-L^{(3,2)}_2}{L^{(3)}_1}, \quad l_{41}=\frac{L^{(4,1)}_1-L^{(3,2)}_3}{L^{(3)}_1}, \\
L_1=\frac{L^{(4)}_3+L^{(4)}_6}{L^{(3)}_1}, \quad L_2=\frac{L^{(4)}_{12}}{L^{(3)}_1}, \quad L_3=\frac{L^{(4)}_{13}}{L^{(3)}_1}, \\
L_4=\frac{L^{(4)}_1+L^{(4)}_{15}}{L^{(3)}_1}, \quad L_5=\frac{L^{(4)}_2+L^{(4)}_{11}}{L^{(3)}_1}, \quad L_6=\frac{L^{(4)}_4+L^{(4)}_{14}}{L^{(3)}_1}, \\
\label{eqn:62}
L_7=\frac{L^{(4)}_5+L^{(4)}_{10}}{L^{(3)}_1}, \quad {L_8=\frac{L^{(4)}_7+L^{(4)}_8}{L^{(3)}_1}}, \quad {L_9=\frac{L^{(4)}_9}{L^{(3)}_1}}
\end{gather}
and $\tau_1$, $\tau_2$ and $\tau_3$ have time dimension.

In the rate equations \eqref{eqn:A} and \eqref{eqn:B} $24$ independent coefficients appear. These equations are the full three-dimensional versions of the one-dimensional equations $(12)$-$(13)$ in \cite{KovVan15a}. They represent the  \textit{generalized ballistic-conductive heat transport laws in three-dimensional isotropic materials}. Equation \eqref{eqn:B} can be rewritten by means three rate equations, splitting the second-order tensor $Q_{ij}$ into its orthogonal components, i.e.
\begin{equation}
\label{eqn:63}
Q_{ij}=Q\delta_{ij}+Q_{\langle ij \rangle}+Q_{[ij]},
\end{equation} 
where 
\begin{align}
\label{Q1}
Q&=\frac{1}{3}Q_{kk} \quad \text{(scalar part of $Q_{ij}$)},\\
\label{Q2}
Q_{\langle ij \rangle}&=\frac{1}{2}(Q_{ij}+Q_{ji})-Q\delta_{ij} \quad \text{(deviator of the symmetric part of $Q_{ij}$)},\\
\label{Q3}
Q_{[ij]}&=\frac{1}{2}(Q_{ij}-Q_{ji}) \quad \text{(skew-symmetric part of $Q_{ij}$)}.
\end{align} 

\noindent From equation \eqref{eqn:B} we derive \textit{the rate equations for} $Q$, $Q_{\langle ij \rangle}$ and $Q_{[ij]}$.

\textit{The rate equation for} $Q$ is ($i=j$)

\begin{equation}
\begin{split}
&3(3\tau_1+\tau_2+\tau_3)\dot{Q}+3(3+l^3_2+l^3_3)Q=(3l_{21}+l_{31}+l_{41})q_{k,k}\\
& \quad +3(L_1+L_2+L_3+3L_9)Q_{,kk}+ (L_4+L_5+L_6+L_7+3L_8)Q_{kl,kl},
\end{split}
\end{equation}
i.e.
\begin{equation}
\label{eqn:B0}
\tau^0\dot{Q}+Q=l^0q_{k,k}+L^0_1Q_{,kk}+L^0_2Q_{kl,kl}\ ,
\end{equation}
where
\begin{gather}
\label{eqn:69}
\tau^0=\frac{3\tau_1+\tau_2+\tau_3}{3+l^3_2+l^3_3}, \quad l^0=\frac{3l_{21}+l_{31}+l_{41}}{3(3+l^3_2+l^3_3)}, \\
L^0_1=\frac{L_1+L_2+L_3+3L_9}{3+l^3_2+l^3_3}, \quad L^0_2=\frac{L_4+L_5+L_6+L_7+3L_8}{3(3+l^3_2+l^3_3)},
\end{gather}
being $\tau^0$ the relaxation time of $Q$;

\textit{the rate equation for $Q_{\langle ij \rangle}$} is
\begin{equation}
\label{eqn:B1}
\overset{\wedge}{\tau}\dot{Q}_{\langle ij \rangle}+Q_{\langle ij \rangle}=\overset{\wedge}{l}q_{\langle i,j \rangle}+{\overset{\wedge}{L}_1Q_{kk,\langle ij \rangle}}+\overset{\wedge}{L}_2Q_{\langle ij \rangle,kk}+\overset{\wedge}{L}_3Q_{k\langle i,j\rangle k}+\overset{\wedge}{L}_4Q_{\langle ik,kj\rangle}\ ,
\end{equation}
where 
\begin{gather}
\label{eqn:72}
\overset{\wedge}{\tau}=\frac{\tau_2+\tau_3}{l^3_2+l^3_3}, \quad \overset{\wedge}{l}=\frac{l_{31}+l_{41}}{l^3_2+l^3_3}, \quad {\overset{\wedge}{L}_1=\frac{L_1}{l^3_2+l^3_3}}, \quad \\
\overset{\wedge}{L}_2=\frac{L_2+L_3}{l^3_2+l^3_3}, \quad
\overset{\wedge}{L}_3=\frac{L_5+L_7}{l^3_2+l^3_3}, \quad \overset{\wedge}{L}_4=\frac{L_4+L_6}{l^3_2+l^3_3},
\end{gather}
being $\overset{\wedge}{\tau}$ the relaxation time of $Q_{\langle ij \rangle}$;

finally \textit{the rate equation for $Q_{[ij]}$} is
\begin{equation}
\label{eqn:B2}
\overset{\vee}{\tau}\dot{Q}_{[ij]}+Q_{[ij]}=\overset{\vee}{l}q_{[i,j]}+\overset{\vee}{L}_1Q_{[ij],kk}+\overset{\vee}{L}_2Q_{k[i,j]k}+\overset{\vee}{L}_3Q_{[ik,kj]}\ ,
\end{equation}
where  
\begin{gather}
\label{eqn:75}
\overset{\vee}{\tau}=\frac{\tau_2-\tau_3}{l^3_2-l^3_3}, \quad \overset{\vee}{l}=\frac{l_{31}-l_{41}}{l^3_2-l^3_3}, \quad \overset{\vee}{L}_1=\frac{L_2-L_3}{l^3_2-l^3_3}, \\
\overset{\vee}{L}_2=\frac{L_7-L_5}{l^3_2-l^3_3}, \quad \overset{\vee}{L}_3=\frac{L_6-L_4}{l^3_2-l^3_3},
\end{gather}
being $\overset{\vee}{\tau}$ the relaxation time of $Q_{[ij]}$.

\section{The rate equations for $q_i$ and $Q_{ij}$ with Onsager reciprocity in the case where $Q_{ij}$ has odd parity}
\label{4.4}
In Section $5$ we have obtained the rate equations for $q_i$ and $Q_{ij}$ and for the scalar part, the deviator of the symmetric part and the skew-symmetric part of $Q_{ij}$ (see \eqref{eqn:A}, \eqref{eqn:B} or \eqref{eqn:A} and \eqref{eqn:B0}, \eqref{eqn:B1}, \eqref{eqn:B2}, respectively) without assuming reciprocity relations, but only isotropy. In this Section we derive the \textit{heat transport laws in three-dimensional isotropic materials} \eqref{eqn:B}, \eqref{eqn:B0}, \eqref{eqn:B1}, \eqref{eqn:B2} in the form  \eqref{eqn:Bbis}, \eqref{eqn:B0bis}, \eqref{eqn:B1bis} and \eqref{eqn:B2bis}, 
by using Onsager reciprocity relations \eqref{eqn:32}-\eqref{eqn:34}, that reduce the number of coefficients in these rate equations from $24$ to $21$ (when compared to the general isotropic case). The phenomenological equations \eqref{eqn:I1} and \eqref{eqn:I2} remain unchanged (thus also the rate equation \eqref{eqn:A}), but equations \eqref{eqn:I3} and \eqref{eqn:I4} assume the following form
\begin{align}
\begin{split}
\label{eqn:I3bis}
B_{kij,k}&=M_1\delta_{ij}\dot{Q}_{kk}+M_2\dot{Q}_{ij}+M_3\dot{Q}_{ji}+{L^{(2,3)}_1}\delta_{ij}q_{k,k}+{L^{(2,3)}_2}q_{i,j} \\
& \quad +{L^{(2,3)}_3}q_{j,i}+L^{(3)}_1\delta_{ij}Q_{kk}+L^{(3)}_2Q_{ij}+L^{(3)}_3Q_{ji}, 
\end{split}
\\
\begin{split}
\label{eqn:I4bis}
B_{ijk}&={L^{(1,4)}_3}\delta_{ij}q_k+{L^{(1,4)}_2}\delta_{ik}q_j+{L^{(1,4)}_1}\delta_{jk}q_i \\
& \quad +\delta_{ij}\left({C^{(4)}_1}Q_{kl,l}+{C^{(4)}_2}Q_{lk,l}+{C^{(4)}_3}Q_{ll,k}\right) \\
& \quad +\delta_{ik}\left({C^{(4)}_4}Q_{jl,l}+{C^{(4)}_5}Q_{lj,l}+{C^{(4)}_2}Q_{ll,j}\right) \\
& \quad +\delta_{jk}\left({C^{(4)}_6}Q_{il,l}+{C^{(4)}_4}Q_{li,l}+{C^{(4)}_1}Q_{ll,i}\right) \\
& \quad +{C^{(4)}_7}Q_{ij,k}+{C^{(4)}_8}Q_{ik,j}+{C^{(4)}_{10}}Q_{jk,i}+{C^{(4)}_{11}}Q_{kj,i} \\
& \quad +{C^{(4)}_9}Q_{ji,k}+{C^{(4)}_{10}}Q_{ki,j},
\end{split}
\end{align}
where, with respect to \eqref{eqn:I3} and \eqref{eqn:I4} 
the coefficients $L^{(3,2)}_i$ have been replaced by $L^{(2,3)}_i$ ($i=1,2,3$), and the coefficients $L^{(4,1)}_i$ by $L^{(1,4)}_i$ ($i=1,2,3$).

\noindent By virtue of \eqref{eqn:I3bis} and \eqref{eqn:I4bis}, (changing $i\rightarrow k$, $j\rightarrow i$, $k\rightarrow j$ in equation \eqref{eqn:I4bis}, deriving it with respect to $x_k$, inserting it into \eqref{eqn:I3bis} and multiplying the obtained equation by $1/L^{(3)}_1$)  we obtain
\begin{equation}
\label{eqn:86_1}
\begin{split}
&\tau_1\delta_{ij}\dot{Q}_{kk}+\tau_2\dot{Q}_{ij}+\tau_3\dot{Q}_{ji}+\delta_{ij}Q_{kk}+l^3_2Q_{ij}+l^3_3Q_{ji}=l_{21}\delta_{ij}q_{k,k}+l_{31}q_{i,j}\\
& \quad +l_{41}q_{j,i}+C_1Q_{kk,ij}+C_2Q_{ij,kk}+C_3Q_{ji,kk}+C_4Q_{jk,ik}+C_5Q_{kj,ik} \\
& \quad +C_6Q_{ik,jk}+C_7Q_{ki,jk}+\delta_{ij}\left(C_8Q_{kl,kl}+C_9Q_{ll,kk}\right),
\end{split}
\end{equation}
where
\begin{gather}
\label{eqn:89}
C_1=\frac{{C^{(4)}_2+C^{(4)}_3}}{L^{(3)}_1}, \quad C_2=\frac{{C^{(4)}_{10}}}{L^{(3)}_1}, \quad C_3=\frac{{C^{(4)}_{11}}}{L^{(3)}_1}, \\
\label{eqn:90}
C_4=\frac{{C^{(4)}_1+C^{(4)}_{10}}}{L^{(3)}_1}, \quad C_5=\frac{{C^{(4)}_2+C^{(4)}_8}}{L^{(3)}_1}, \quad C_6=\frac{{C^{(4)}_4+C^{(4)}_9}}{L^{(3)}_1}, \\
\label{eqn:91}
C_7=\frac{{C^{(4)}_5+C^{(4)}_7}}{L^{(3)}_1}, \quad C_8=\frac{{C^{(4)}_4+C^{(4)}_6}}{L^{(3)}_1}, \quad C_9=\frac{{C^{(4)}_1}}{L^{(3)}_1}.
\end{gather}
The rate equation \eqref{eqn:86_1} is the same as \eqref{eqn:B}, but with $L_i$ replaced by $C_i$ ($i=1\ldots 9$).

We remark that the coefficients $l_{21}$, $l_{31}$ and $l_{41}$ in \eqref{eqn:86_1} transform according to Onsager relations \eqref{eqn:32} and \eqref{eqn:33}, so that we have 
\begin{equation}
\label{l1234}
l_{21}=\left(L^{(1,4)}_1-L^{(2,3)}_1\right)/L^{(3)}_1, \quad l_{31}=\left(L^{(1,4)}_2-L^{(2,3)}_2\right)/L^{(3)}_1, \quad l_{41}=\left(L^{(1,4)}_3-L^{(2,3)}_3\right)L^{(3)}_1.
\end{equation}
By virtue of \eqref{l1234} and $\eqref{eqn:89}_2$, $\eqref{eqn:90}_1$ and $\eqref{eqn:91}_3$, the further conditions for the coefficients are worked out:
\begin{equation}
\label{79}
l_{31}=-L^{(1)}l_{14}/L^{(3)}_1, \quad l_{41}=-L^{(1)}(l_{12}+l_{13})/L^{(3)}_1-l_{21}, \quad C_9=C_4-C_2.
\end{equation}

\noindent Thus, using relations \eqref{79}, the rate equation \eqref{eqn:86_1} for $Q_{ij}$ takes the form
\begin{equation}
\label{eqn:Bbis}
\begin{split}
&\tau_1\delta_{ij}\dot{Q}_{kk}+\tau_2\dot{Q}_{ij}+\tau_3\dot{Q}_{ji}+\delta_{ij}Q_{kk}+l^3_2Q_{ij}+l^3_3Q_{ji}=l_{21}\delta_{ij}q_{k,k}-L^{(1)}l_{14}/L^{(3)}_1 q_{i,j} \\
& \quad -\left[L^{(1)}(l_{12}+l_{13})/L^{(3)}_1+l_{21}\right]q_{j,i}+C_1Q_{kk,ij}+C_2Q_{ij,kk}+C_3Q_{ji,kk}+C_4Q_{jk,ik}+C_5Q_{kj,ik} \\
& \quad +C_6Q_{ik,jk}+C_7Q_{ki,jk}+\delta_{ij}\left[C_8Q_{kl,kl}+({C_4-C_2})Q_{ll,kk}\right].
\end{split}
\end{equation}

As in \eqref{eqn:63}, we split the second-order tensor $Q_{ij}$ in its orthogonal components  $Q$, $Q_{\langle ij \rangle}$ and $Q_{[ij]}$, its scalar part,  the deviator of its symmetric part, its skew-symmetric part (see \eqref{Q1}-\eqref{Q3}) that, having $Q_{ij}$ odd parity, have also odd parity. In the following we work out the rate equations for  $Q$, $Q_{\langle ij \rangle}$ and $Q_{[ij]}$. 

Thus, from equation \eqref{eqn:Bbis} we derive:

\medskip

\textit{the rate equation for $Q$} (obtained when $i=j$) 

\begin{equation}
\label{eqn:B0bis}
\tau^0\dot{Q}+Q=c^0q_{k,k}+C^0_1Q_{,kk}+C^0_2Q_{kl,kl}\ ,
\end{equation}
where $\tau^0$ is given by $\eqref{eqn:69}_1$ and
\begin{equation}
\label{c0}
c^0=\frac{2L^{(3)}_1l_{21}-L^{(1)}(l_{12}+l_{13}+l_{14})}{3L^{(3)}_1(3+l^3_2+l^3_3)}, \quad  C^0_1=\frac{C_1{-2C_2}+C_3+{3C_4}}{3+l^3_2+l^3_3}, \quad C^0_2=\frac{C_4+C_5+C_6+C_7+3C_8}{3(3+l^3_2+l^3_3)};
\end{equation}

\textit{the rate equation for $Q_{\langle ij \rangle}$} 
\begin{equation}
\label{eqn:B1bis}
\overset{\wedge}{\tau}\dot{Q}_{\langle ij \rangle}+Q_{\langle ij \rangle}=\overset{\wedge}{c}q_{\langle i,j \rangle}+\overset{\wedge}{C}_1Q_{kk,\langle ij \rangle}+\overset{\wedge}{C}_2Q_{\langle ij \rangle,kk}+\overset{\wedge}{C}_3Q_{k\langle i,j\rangle k}+\overset{\wedge}{C}_4Q_{\langle ik,kj\rangle}\ ,
\end{equation}
where $\overset{\wedge}{\tau}$ is given by $\eqref{eqn:72}_{1}$ and 
\begin{equation}
\label{cc}
\overset{\wedge}{c}=-\frac{L^{(3)}_1l_{21}+L^{(1)}(l_{12}+l_{13}+l_{14})}{L^{(3)}_1(l^3_2+l^3_3)}, \;\;\;  
\overset{\wedge}{C}_1=\frac{C_1}{l^3_2+l^3_3}, \;\;\;  
\overset{\wedge}{C}_2=\frac{C_2+C_3}{l^3_2+l^3_3}, \;\;\; 
\overset{\wedge}{C}_3=\frac{C_5+C_7}{l^3_2+l^3_3}, \;\;\; \overset{\wedge}{C}_4=\frac{C_4+C_6}{l^3_2+l^3_3};
\end{equation}

\textit{the rate equation for $Q_{[ij]}$}
\begin{equation}
\label{eqn:B2bis}
\overset{\vee}{\tau}\dot{Q}_{[ij]}+Q_{[ij]}=\overset{\vee}{c}q_{[i,j]}+\overset{\vee}{C}_1Q_{[ij],kk}+\overset{\vee}{C}_2Q_{k[i,j]k}+\overset{\vee}{C}_3Q_{[ik,kj]}\ ,
\end{equation}
where $\overset{\vee}{\tau}$ is given by $\eqref{eqn:75}_{1}$ and  
\begin{equation}
\label{ccc}
\overset{\vee}{c}=\frac{L^{(3)}_1l_{21}+L^{(1)}(l_{12}+l_{13}-l_{14})}{L^{(3)}_1(l^3_2-l^3_3)}, \quad 
\overset{\vee}{C}_1=\frac{C_2-C_3}{l^3_2-l^3_3}, \quad
\overset{\vee}{C}_2=\frac{C_7-C_5}{l^3_2-l^3_3}, \quad \overset{\vee}{C}_3=\frac{C_6-C_4}{l^3_2-l^3_3}.
\end{equation}

\subsection{One-dimensional heat transport in the case where $Q_{ij}$ has odd parity}
In this Subsection we focus on the one-dimensional case, in order to appreciate how the generalization from one dimension to three dimensions analysed in this paper is far from trivial. In the one-dimensional case we have that the components of  \textbf{B} and \textbf{Q} reduce to 
\begin{equation}
\label{87}
B\equiv B_{111} \quad \text{and}\quad  Q=Q_{11}, \quad \text{and}\quad
\mathbf{q}=(q,0,0), \quad 
\mathbf{Q}=
\begin{pmatrix}
Q & 0 & 0 \\
0 & 0 & 0 \\
0 & 0 & 0
\end{pmatrix}
, \quad 
\mathbf{b}=
\begin{pmatrix}
b & 0 & 0 \\
0 & 0 & 0 \\
0 & 0 & 0
\end{pmatrix}.
\end{equation}
The system  of equations \eqref{eqn:I1}-\eqref{eqn:I4} (in which we use the Onsager relations assuming that $Q_{ij}$ has odd parity) becomes 
\begin{align}
\label{eqn:mon_1}
m\dot{q}-b_{,x}&=-L^{(1)}q-L^{(1,4)}Q_{,x}, \\
\label{eqn:mon_2}
b-\frac{1}{T}&=L^{(2)}q_{,x}+L^{(2,3)}Q, \\
\label{eqn:mon_3}
M\dot{Q}-B_{,x}&=-L^{(2,3)}q_{,x}-L^{(3)}Q, \\
\label{eqn:mon_4}
B&=L^{(1,4)}q+C^{(4)}Q_{,x},
\end{align}
where $m>0$, $M>0$ (see \cite{KovVan15a}) and
\begin{gather}
L^{(1,4)}=L^{(1,4)}_1+L^{(1,4)}_2+L^{(1,4)}_3, \quad L^{(2)}=L^{(2)}_1+L^{(2)}_2+L^{(2)}_3, \\[0.5em]
 L^{(2,3)}=L^{(2,3)}_1+L^{(2,3)}_2+L^{(2,3)}_3, \quad M=M_1+M_2+M_3, \\[0.5em]
L^{(3)}=L^{(3)}_1+L^{(3)}_2+L^{(3)}_3, \\[0.5em] 
\begin{split}
C^{(4)}=2C^{(4)}_1+2C^{(4)}_2+C^{(4)}_3+2C^{(4)}_4&+C^{(4)}_5+C^{(4)}_6+C^{(4)}_7+C^{(4)}_8+ \\ 
 & +C^{(4)}_9+2C^{(4)}_{10}+C^{(4)}_{11}, 
\end{split}
\end{gather}
with  $(\cdot)_{,x}$ indicating the derivative of $(\cdot)$ with respect to  $x$. We observe that the system of equations \eqref{eqn:mon_1}-\eqref{eqn:mon_4} obtained here is more general of equations $(7)$-$(10)$ deduced in \cite{KovVan15a}, because of the presence of the phenomenological constant $L^{(1,4)}$ in \eqref{eqn:mon_1} and \eqref{eqn:mon_4} and the fact that the coefficients  $L^{(1,4)}$,  $L^{(2)}$,  $L^{(2,3)}$, $M$,  $L^{(3)}$,  $C^{(4)}$ have been obtained from a three-dimensional approach.

In this case, the entropy production \eqref{eqn:EI} assumes the form

\begin{equation}
\label{eqn:sigma_mon}
\sigma^{(s)}=L^{(1)}q^2+L^{(2)}(q_{,x})^2+L^{(3)}Q^2+C^{(4)}(Q_{,x})^2 + 2L^{(1,4)}qQ_{,x}+2L^{(2,3)}q_{,x}Q\geq 0,
\end{equation}
with $C^{(4)}=L_{111111}$ (see matrix \eqref{eqn:BigMatrix} of the Appendix), or in  symbolic matrix notation
\begin{equation}
\label{97}
\sigma^{(s)}=
\begin{pmatrix}
q & q_{,x} & Q & Q_{,x}
\end{pmatrix}
\underbrace{
\begin{pmatrix}
L^{(1)} & 0 & 0 & L^{(1,4)} \\[0.5em]
0 & L^{(2)} & L^{(2,3)} & 0 \\[0.5em]
0 & L^{(2,3)} & L^{(3)} & 0 \\[0.5em]
L^{(1,4)} & 0 & 0 & C^{(4)}
\end{pmatrix}
}_{\displaystyle \mathcal{A}}
\begin{pmatrix}
q \\[0.5em] 
q_{,x} \\[0.5em] 
Q \\[0.5em] 
Q_{,x}
\end{pmatrix}
\geq 0.
\end{equation}

\noindent Because the bilinear  form \eqref{eqn:sigma_mon} must be non-negative, the matrix $\mathcal{A}$ (that is symmetric) associated to this form  is non-negative semi-definite, so that the elements of its main diagonal and its  major minors must be non-negative

\begin{gather}
\label{98}
L^{(1)}\geq 0, \quad L^{(2)}\geq 0, \quad L^{(3)}\geq 0, \quad C^{(4)}\geq 0, \\[0.5em]
\label{99}
L^{(2)}L^{(3)}-\left(L^{(2,3)}\right)^2\geq 0, \quad L^{(1)}C^{(4)}-\left(L^{(1,4)}\right)^2\geq 0.
\end{gather}

Using \eqref{eqn:mon_2} and \eqref{eqn:mon_4}, equations \eqref{eqn:mon_1} and \eqref{eqn:mon_3} become

\begin{align}
\label{eqn:125}
mq_{,t}+L^{(1)}q-L^{(2)}q_{,xx}&=\left(\frac{1}{T}\right)_{,x}-DQ_{,x},\\
\label{eqn:126}
MQ_{,t}+L^{(3)}Q-C^{(4)}Q_{,xx}&=Dq_{,x},
\end{align}

\noindent where $D=L^{(1,4)}-L^{(2,3)}$. 

\noindent In the following we introduce the relaxation time of the internal variable $Q$, called $\tau^J$:
\begin{equation}
\label{tauJ}
\tau^J=\frac{M}{L^{(3)}}.
\end{equation}

\noindent Furthermore, we have supposed the body is at rest, so that material derivative coincides with the partial time derivative $(\cdot)_{,t}$. 

Equations \eqref{eqn:125} and \eqref{eqn:126} are analogous  to equations $(12)$ and $(13)$  of  \cite{KovVan15a}. For $L^{(2)}=C^{(4)}=0$, these equations coincide with those provided in \cite{JouAta92b} or \cite{SelEta16b} by assuming $Q_{ij}$ as the flux of the heat flux.

In the following we will derive heat transport equations analogous but more general of that obtained in \cite{KovVan15a}, where the finite speed of thermal disturbances and the ballistic and diffusive motion of phonons (heat carriers) are taken into account. Instead, in Fourier equation the velocity of heat propagation is infinite.
Differentiating equation \eqref{eqn:125} with respect to time, equation \eqref{eqn:126} with respect to the spatial variable $x$ and using equation \eqref{eqn:125} and its second spatial derivative, we can eliminate $Q$ and work out the following \textit{generalized ballistic-conductive  heat transport law} 
\begin{equation}
\label{eqn:heat}
\begin{split}
&mMq_{,tt}+\left(ML^{(1)}+mL^{(3)}\right)q_{,t}-\left(mC^{(4)}+ML^{(2)}\right)q_{,xxt}+C^{(4)}L^{(2)}q_{,xxxx} \\
& \quad -\left(L^{(1)}C^{(4)}+H\right)q_{,xx}+L^{(3)}L^{(1)}q=M\left(\frac{1}{T}\right)_{,xt}+L^{(3)}\left(\frac{1}{T}\right)_{,x} -C^{(4)}\left(\frac{1}{T}\right)_{,xxx},
\end{split}
\end{equation}
where 
\begin{equation}
\label{H}
H=L^{(3)}L^{(2)}-D^2.
\end{equation} 
Equation \eqref{eqn:heat} has been obtained via several differentiations of the linear governing equations \eqref{eqn:125} and \eqref{eqn:126}. Hence, equation \eqref{eqn:heat} is not equivalent to the system of equations \eqref{eqn:125} and \eqref{eqn:126}. In fact \eqref{eqn:heat} has a larger set of solutions, coming from the larger number of necessary initial conditions.

\noindent Thus, we derive 
\begin{equation}
\label{eqn:heat_1}
\tau\tau^Jq_{,tt}+\tau^qq_{,t}+q-\alpha q_{,xxt}+\beta q_{,xxxx}-\gamma q_{,xx}=\nu\left(\frac{1}{T}\right)_{,xt}-\lambda T_{,x} -\zeta\left(\frac{1}{T}\right)_{,xxx},
\end{equation}
where 
\begin{gather}
\label{105}
\tau^q=\tau+\tau^J, \quad \nu=\frac{M}{L^{(1)}L^{(3)}},   \\
\label{106}
 \gamma=\frac{L^{(1)}C^{(4)}+H}{L^{(1)}L^{(3)}}, \quad \beta=\frac{C^{(4)}L^{(2)}}{L^{(1)}L^{(3)}}, \quad \alpha=\frac{mC^{(4)}+ML^{(2)}}{L^{(1)}L^{(3)}}, \quad \zeta=\frac{C^{(4)}}{L^{(1)}L^{(3)}},
\end{gather}
$H$ is defined by \eqref{H} and $\lambda$ is given by \eqref{eqn:54}$_2$.

\noindent In \eqref{eqn:heat_1} wee see that relaxation time $\tau^q=\tau+\tau^J$ is given by two contributions: the first comes from the relaxation time of the heat flux (see \eqref{eqn:54}$_1$) and the second comes from the relaxation time of the internal variable (see \eqref{tauJ}).  

\subsection{Special cases of heat transport equation in the assumption that $Q_{ij}$ has odd parity}
From \eqref{eqn:heat}, it is possible to derive as particular case some special equations which have been often analysed in the literature on heat transport. 

\medskip

\paragraph{\textit{Ballistic-conductive equation}.} In the case where $C^{(4)}=L^{(2)}=0$, the heat equation \eqref{eqn:heat} becomes
\begin{equation}
\label{107}
mMq_{,tt}+\left(ML^{(1)}+mL^{(3)}\right)q_{,t}-D^2q_{,xx}+L^{(3)}L^{(1)}q=M\left(\frac{1}{T}\right)_{,xt}+L^{(3)}\left(\frac{1}{T}\right)_{,x}.
\end{equation}
Thus, we can write
\begin{equation}
\label{108}
\tau\tau^Jq_{,tt}+\tau^qq_{,t}+q-\eta q_{,xx}=\nu\left(\frac{1}{T}\right)_{,xt}-\lambda T_{,x},
\end{equation}
where 
\begin{equation}
\label{eta}
\eta=\frac{D^2}{L^{(1)}L^{(3)}}.
\end{equation}

\medskip

\paragraph{\textit{Guyer-Krumhansl equation}.}
In the case where $C^{(4)}=M=0$, the heat equation \eqref{eqn:heat} becomes
\begin{equation}
\label{109}
mL^{(3)}q_{,t}-Hq_{,xx}+L^{(3)}L^{(1)}q=L^{(3)}\left(\frac{1}{T}\right)_{,x},
\end{equation}
then, we work out
\begin{equation}
\label{110}
\tau q_{,t}-l^2 q_{,xx}+q=-\lambda T_{,x},
\end{equation}
with
\begin{equation}
\label{111}
l^2=\frac{H}{L^{(1)}L^{(3)}},
\end{equation}
where $l$, having the dimension of a length which may be interpreted as an average, mean free path of the heat carriers  (phonons) i.e. the average length between successive collision amongst them. We observe that only in Guyer-Krumhansl heat equation the coefficient multiplying the field $q_{,xx}$ has the physical meaning of $l^2$.

\medskip

\paragraph{\textit{Cahn-Hilliard type equation}.}
In the case where $C^{(4)}=M=m=0$, the heat equation \eqref{eqn:heat} becomes
\begin{equation}
\label{112}
L^{(3)}L^{(1)}q-Hq_{,xx}=L^{(3)}\left(\frac{1}{T}\right)_{,x},
\end{equation}
from which we obtain
\begin{equation}
\label{113}
q-l^2 q_{,xx}=-\lambda T_{,x}.
\end{equation}

\medskip

\paragraph{\textit{Jeffreys type equation} (or double-lag model \cite{Tzou}).}
In the case where $C^{(4)}=L^{(2)}=m=D=0$ (then, $H=0$), the heat equation \eqref{eqn:heat} becomes
\begin{equation}
\label{114}
ML^{(1)}q_{,t}+L^{(3)}L^{(1)}q=M\left(\frac{1}{T}\right)_{,xt}+L^{(3)}\left(\frac{1}{T}\right)_{,x},
\end{equation}
thus we derive:
\begin{equation}
\label{115}
\tau^Jq_{,t}+q=\nu\left(\frac{1}{T}\right)_{,xt}-\lambda T_{,x}.
\end{equation}
We note that in the Jeffreys type heat equation $\tau^J$ is the relaxation time of $q$.

\medskip

\paragraph{\textit{Maxwell-Cattaneo-Vernotte equation}.}
In the case where $C^{(4)}=M=L^{(2)}=D=0$ (then, $H=0$), the heat equation \eqref{eqn:heat} becomes
\begin{equation}
\label{116}
mq_{,t}+L^{(1)}q=\left(\frac{1}{T}\right)_{,x},
\end{equation}
from which we have:
\begin{equation}
\label{117}
\tau q_{,t}+q=-\lambda T_{,x}.
\end{equation}

\medskip

\paragraph{\textit{Fourier equation}.}
In the case where $C^{(4)}=M=L^{(2)}=D=m=0$ (then, $H=0$), the heat equation \eqref{eqn:heat} becomes
\begin{equation}
\label{118}
L^{(1)}q=\left(\frac{1}{T}\right)_{,x},
\end{equation}
i.e.
\begin{equation}
\label{119}
q=-\lambda T_{,x}.
\end{equation}

What is specially worth in this Subsection is not only the ability to obtain many situations studied up to now, but specially the fact that the coefficients appearing in the one-dimensional case are complicated combinations of the independent coefficients appearing in the three-dimensional case. Thus, measurements in one dimension are not sufficient to give information in the general three-dimensional situation, which is the only one able to exhibit the basic meaning of each coefficient. We emphasize that Jeffrey type, Maxwell-Cattaneo-Vernotte and Fourier equations are the same as in \cite{KovVan15a}.


\section{Rate equations for $q_i$ and $Q_{ij}$ in the isotropic case where $Q_{ij}$ has even parity}

In Section $5$ we have obtained the rate equations \eqref{eqn:A} and \eqref{eqn:B} for the heat flux $q_i$ and the internal variable $Q_{ij}$, respectively, in the general isotropic case without assumptions regarding the parity of the internal variable $Q_{ij}$ ($q_i$ is odd and $Q_{ij}$ can be of odd or even type) and than we have not discussed Onsager reciprocity relations. In Section $6$ we have shown  how these rate equations transform supposing the odd parity of $Q_{ij}$. In this Section we treat the case where $Q_{ij}$ has even parity and it is very easy to see that the rate equation \eqref{eqn:A} remains unchanged (as in the odd parity case). Instead, the rate equation \eqref{eqn:B} (that takes the form \eqref{eqn:Bbis} when we assume the even parity of $Q_{ij}$) transforms in 
\begin{equation}
\label{eqn:BbisE}
\begin{split}
&\tau_1\delta_{ij}\dot{Q}_{kk}+\tau_2\dot{Q}_{ij}+\tau_3\dot{Q}_{ji}+\delta_{ij}Q_{kk}+l^3_2Q_{ij}+l^3_3Q_{ji}=l_{21}\delta_{ij}q_{k,k}+L^{(1)}l_{14}/L^{(3)}_1q_{i,j}\\
& \quad +\left[L^{(1)}(l_{12}+l_{13})/L^{(3)}_1-l_{21}\right]q_{j,i}+C_1Q_{kk,ij}+C_2Q_{ij,kk}+C_3Q_{ji,kk}+C_4Q_{jk,ik}+C_5Q_{kj,ik} \\
& \quad +C_6Q_{ik,jk}+C_7Q_{ki,jk}+\delta_{ij}\left[C_8Q_{kl,kl}+({C_4-C_2})Q_{ll,kk}\right],
\end{split}
\end{equation}
where the quantities $l_{21}$, $l_{31}$ and $l_{41}$ take the following form
\begin{equation}
\label{lE}
l_{21}=\frac{L^{(2,3)}_1-L^{(1,4)}_1}{L^{(3)}_1}, \quad l_{31}=\frac{L^{(2,3)}_2-L^{(1,4)}_2}{L^{(3)}_1}, \quad l_{41}=\frac{L^{(2,3)}_3-L^{(1,4)}_3}{L^{(3)}_1},
\end{equation}
in which Onsager symmetry relations \eqref{eqn:32} and \eqref{eqn:33} have been applied, with negative sign. The other coefficients continue to have the same definitions given in Section $5$, but the coefficients of $q_{i,j}$ and $q_{j,i}$ have different signs  with respect to those in \eqref{eqn:Bbis}.
Furthermore, in this considered case relations \eqref{79}$_{1,2}$ become
\begin{equation}
\label{79E}
l_{31}=L^{(1)}l_{14}/L^{(3)}_1, \quad l_{41}=L^{(1)}(l_{12}+l_{13})/L^{(3)}_1-l_{21}.
\end{equation}

Finally, the rate equations for the orthogonal components $Q$, $Q_{\langle ij \rangle}$ and $Q_{[ij]}$, that are still of even type, remain formally unchanged from the equations \eqref{eqn:B0bis}, \eqref{eqn:B1bis}, and \eqref{eqn:B2bis}, valid when $Q_{ij}$ is of odd type. But we have to emphasize that the quantities $c^0$, $\overset{\wedge}{c}$ and $\overset{\vee}{c}$ (defined by \eqref{c0}$_1$, \eqref{cc}$_1$ and \eqref{ccc}$_1$, respectively) take the following different form
\begin{gather}
\label{cE}
c^0=\frac{2L^{(3)}_1l_{21}+L^{(1)}(l_{12}+l_{13}+l_{14})}{3L^{(3)}_1(3+l^3_2+l^3_3)}, \qquad \overset{\wedge}{c}=\frac{L^{(1)}(l_{12}+l_{13}+l_{14})-L^{(3)}_1l_{21}}{L^{(3)}_1(l^3_2+l^3_3)}, \\[0.5em]
\overset{\vee}{c}=\frac{L^{(3)}_1l_{21}-L^{(1)}(l_{12}+l_{13}+l_{14})}{L^{(3)}_1(l^3_2-l^3_3)}.
\end{gather}

\subsection{One-dimensional isotropic heat transport in the assumption that $Q_{ij}$ has even parity}
Taking into account expressions \eqref{87}, using the Onsager relations \eqref{eqn:O-C1}-\eqref{eqn:O-C5} in the case where the internal variable $Q_{ij}$ has an even parity, equations \eqref{eqn:I1}-\eqref{eqn:I4} take the form 
\begin{align}
\label{eqn:mon_1E}
m\dot{q}-b_{,x}&=-L^{(1)}q-L^{(1,4)}Q_{,x}, \\
\label{eqn:mon_2E}
b-\frac{1}{T}&=L^{(2)}q_{,x}+L^{(2,3)}Q, \\
\label{eqn:mon_3E}
M\dot{Q}-B_{,x}&=L^{(2,3)}q_{,x}-L^{(3)}Q, \\
\label{eqn:mon_4E}
B&=-L^{(1,4)}q+C^{(4)}Q_{,x},
\end{align}
where only equations \eqref{eqn:mon_3E} and \eqref{eqn:mon_4E} are different from \eqref{eqn:mon_3} and \eqref{eqn:mon_4} because of the signs of the first terms in their right-hand sides. As consequence of this difference we have that the entropy production \eqref{eqn:sigma_mon} takes the new reduced form
\begin{equation}
\label{eqn:sigma_monE}
\sigma^{(s)}=L^{(1)}q^2+L^{(2)}(q_{,x})^2+L^{(3)}Q^2+C^{(4)}(Q_{,x})^2,
\end{equation}
so that the associated matrix $\mathcal{A}$ (that is diagonal and thus symmetric) takes the following diagonal form 
\begin{equation}
\mathcal{A}=
\begin{pmatrix}
L^{(1)} & 0 & 0 & 0 \\[0.5em]
0 & L^{(2)} & 0 & 0 \\[0.5em]
0 & 0 & L^{(3)} & 0 \\[0.5em]
0 & 0 & 0 & C^{(4)}
\end{pmatrix},
\end{equation}
so that only relations \eqref{98} are still true.

\noindent Furthermore, using \eqref{eqn:mon_2E} and \eqref{eqn:mon_4E}, equations \eqref{eqn:mon_1E} and \eqref{eqn:mon_3E} become
\begin{align}
\label{eqn:125E}
mq_{,t}+L^{(1)}q-L^{(2)}q_{,xx}&=\left(\frac{1}{T}\right)_{,x}-DQ_{,x},\\
\label{eqn:126E}
MQ_{,t}+L^{(3)}Q-C^{(4)}Q_{,xx}&=-Dq_{,x},
\end{align}
where $D=L^{(1,4)}-L^{(2,3)}$.
Relation \eqref{eqn:125E} is equal to \eqref{eqn:125}, while \eqref{eqn:126E} has opposite sign in its right-hand side with respect to \eqref{eqn:126} ($D$ continues to have the same value).

\noindent Finally, deriving equation \eqref{eqn:125E} with respect to time, equation \eqref{eqn:126E} with respect to the spatial variable $x$ and using equation \eqref{eqn:125E} and its second spatial derivative, we can eliminate $Q$ and work out the same the heat transport equation \eqref{eqn:heat} (and than \eqref{eqn:heat_1}) where the only difference consist in the fact that the quantity $H$ defined by \eqref{H} takes the new form
\begin{equation}
\label{HE}
H=L^{(3)}L^{(2)}+D^2.
\end{equation}
Thus, $H$ is always positive in the case of even parity of $Q_{ij}$.

\subsection{Special cases of heat transport equation in the assumption that $Q_{ij}$ has even parity}

Applying the procedures used in Subsection $6.2$ to obtain from \eqref{eqn:heat} (and also \eqref{eqn:heat_1}) special cases, we derive the following results:

\medskip

\noindent \textit{a) The Ballistic-Conductive heat transport equations} \eqref{107} and \eqref{108}, being $C^{(4)}=L^{(2)}=0$ and $H=L^{(3)}L^{(2)}+D^2$, take the following form
\begin{gather}
\label{143}
mMq_{,tt}+\left(ML^{(1)}+mL^{(3)}\right)q_{,t}+D^2q_{,xx}+L^{(3)}L^{(1)}q=M\left(\frac{1}{T}\right)_{,xt}+L^{(3)}\left(\frac{1}{T}\right)_{,x}, \\[0.5em]
\label{144}
\tau\tau^Jq_{,tt}+\tau^qq_{,t}+q+\eta q_{,xx}=\nu\left(\frac{1}{T}\right)_{,xt}-\lambda T_{,x},
\end{gather}
different from \eqref{107} and \eqref{108} because of the plus sign before $D^2$ and $\eta$. In \eqref{143} and \eqref{144} the definitions  \eqref{eqn:54}$_{1,2}$, \eqref{tauJ}, \eqref{105} and \eqref{eta} and  are still valid; 

\medskip

\noindent \textit{b) Guyer-Krumhansl heat transport equations}, being $C^{(4)}=M=0$, have the expressions same as \eqref{109} and \eqref{110}, but with $H$ (see \eqref{111}) replaced by  \eqref{HE};

\medskip

\noindent \textit{c) Cahn-Hilliard type heat transport equations} \eqref{112} and \eqref{113}, being $C^{(4)}=M=m=0$, remain unchanged, but  with $H$ replaced by  \eqref{HE}; 

\medskip

\noindent \textit{d) Jeffreys type, Maxwell-Cattaneo-Vernotte, Fourier heat transport equations}  (where $H$ is not present), remain unchanged.

\section{Discussion and conclusions}

In this paper, ballistic-conductive heat transport in isotropic materials has been treated in the framework of Non-Equilibrium Thermodynamics with Internal Variables (NET-IV). Onsager reciprocity has also been considered (in both particular cases in which $Q_{ij}$ have been assumed to be odd or even with respect to macrocopic time reversal) and the consequences were derived. For the sake of fast applicability the explicit expressions for the components of the conductive matrix are given in the Appendix in the two cases.\footnote{Remarkable, that for nonlinear, or quasilinear conductivity tensors one can get more restrictions (see \cite{K2} and \cite{K3}).}

Our approach, NET-IV, is general and universal. It characterises the deviation from local equilibrium both in the entropy density and in the entropy flux in the simplest possible functional forms.  The entropy density depends on the internal variables quadratically, in order to preserve the concavity, that is thermodynamic stability. The entropy flux depends on the internal variables linearly therefore it disappears when they are zero. As long as these two physical conditions and the entropy inequality are valid, the derived consequences are also valid. The generality of the assumptions ensure the universality of the final evolution equations. Here we have considered a strictly  linear theory, when the \textbf{m} and \textbf{M} tensors and the conductivity tensors, $\textbf{L}^{(\alpha,\beta)}$ and $\textbf{L}^{(\gamma)}$, are constant.

The conditions of positive definiteness of the corresponding conductivity matrix can be calculated directly with the help of computer algebra programs. Though the expressions are very cumbersome, it should be noted that every coefficient appearing in them corresponds in principle to an observable phenomenon. Instead, the much simpler one-dimensional case may grasp essential qualitative features, but its coefficients are a combination of three-dimensional coefficients giving a deeper and more complete description.
We have obtained a complete set of equations for generalized ballistic-conductive heat transport in three-dimensional isotropic rigid conductors for the variables $T,q_i,Q_{ij}$. These are the balance of internal energy \eqref{balance1} with the caloric equation of state $s'_{eq}(e)=1/T$ and the balance type constitutive equations \eqref{eqn:A}, \eqref{eqn:B} (or  \eqref{eqn:A}, \eqref{eqn:B0}, \eqref{eqn:B1}, \eqref{eqn:B2}) in the general isotropic case, and \eqref{eqn:A}, \eqref{eqn:Bbis} (or \eqref{eqn:A}, \eqref{eqn:B0bis}, \eqref{eqn:B1bis},  \eqref{eqn:B2bis}) with Onsagerian reciprocity as additional constraints. 

There are two different aspects of ballistic heat transport in continua. From the point of view of kinetic theory it is the propagation of phonons without collisions with the lattice. Then heat is reflected only at the boundaries of the medium. This microscopic understanding is the foundation of the so-called ballistic-diffusive integrodifferential model of Chen \cite{Che01a, Che02a, TanEta16a, TanEta16a1, TanEta17a}. There kinetic theory and macroscopic considerations are mixed, the distribution function $f$ is split into two parts, one for ballistic phonons and the other referred to diffusive phonons. Also, internal energy and heat flux are decomposed into ballistic and diffusive components. This approach leads to two independent continuum representations. First, it is a  particular boundary condition for continuum theories that can also be introduced to second-sound models, like Guyer-Krunhansl equation \cite{AlvEta12a}. On the other hand, for ballistic phonons, the speed of propagation is equal to the speed of 'first' sound, the speed of elastic waves in the medium. The speed of propagation is independent of the boundary conditions in a continuum approach, and this is the meaning of the ballistic terminology in our theory, following Rational Extended Thermodynamics (RET) \cite{DreStr93a, MulRug98b}.  It is also remarkable that Chen's model is equivalent to an extended continuum heat transport theory, where the coexistence of two kinds of heat carriers (ballistic and diffusive phonons) is assumed as it was shown by Lebon et al. \cite{LebEta11a, Leb14a} and investigated in \cite{VazEta20a}.

Theories of Extended Thermodynamics (ET) assume that the constitutive equations are local, and the rate equations are written in a hierarchical series of balances, where the dissipative fluxes appear as densities in the consecutive balance. These assumptions are consequences of the definition of the macroscopic fields as moments of the single-particle phase space probability density and the Boltzmann equation. In our case, with internal variables, this structure is the consequence of the second law and can be observed on the left-hand side of (\ref{eqn:general1}) and (\ref{eqn:general3}). Then essential aspects of ET are well represented. On the other hand, NET-IV has many material coefficients that are missing in ET, in particular in Rational Extended Thermodynamics, where only the two relaxation times of the Callaway collision integral represent the material properties. This property of RET is attractive, but the price is not only that the validity of the theory is connected to the particularities of the microscopic model, but also that the speed of the ballistic propagation, the speed of elastic waves, can be obtained exactly only by considering the complete moment series, or practically by using dozens of evolution equations (with consecutively increasing tensorial orders) \cite{MulRug98b}. The low number of material coefficients leads to many evolution equations in modelling ballistic propagation of heat.

Giving the three-dimensional structure of ET and NET-IV for heat transport in case of isotropic materials opens the field to build and solve realistic models of two- and three-dimensional experimental setups, where the two theories lead to different predictions.  
To appreciate some of the original aspects of this work, let us eventually comment the equations \eqref{eqn:I2} and \eqref{eqn:I4} for $b_{ij}$ and $B_{ijk}$ and their consequences on the entropy flux. It is well known in the literature \cite{SelEta16b, SelEta13} that one of the expression of the entropy flux is 
\begin{equation}
\label{c}
J_i=\frac{1}{T}q_i+\frac{l^2}{\lambda T^2}q_{i,j}q_j,
\end{equation}

Note then that the constitutive equation for the entropy flux \eqref{eqn:3}, when $b_{ij}$ and $B_{ijk}$ are given by \eqref{eqn:I2} and \eqref{eqn:I4},  lead to a richer expression than \eqref{c}, namely
\begin{equation}
\begin{split}
J_i&=\frac{1}{T} q_i+L^{(2)}_1q_iq_{k,k}+L^{(2)}_2q_{i,j}q_j+L^{(2)}_3q_jq_{j,i}+L^{(2,3)}_1q_iQ_{kk} +L^{(2,3)}_2Q_{ij}q_j+L^{(2,3)}_3q_jQ_{ji} \\
& \quad +f(L^{(4,1)}_i,L^{(4)}_i,q_i,Q_{ij,k}).
\end{split}
\end{equation}
Thus, in our analysis the extended entropy flux is more general than (\ref{c}), and plays an important role in the thermodynamic consistency of couplings with the heat flux $q_i$ and tensorial internal variables as $Q_{ij}$.

\section{Acknowledgement}   
The work was supported by the grants National Research, Development and Innovation Office: NKFIH 116197(116375), NKFIH 124366(124508) and NKFIH 123815. The authors thank Prof.  David Jou, from  Universitat  Auton\`oma di Barcelona, Catalonia,   Spain, for his appreciated comments and remarks,  and  Robert Kov\'acs, from BME, Hungary,  for his valuable discussions. The insightful comments of our second referee were also welcome.

\newpage
\section*{Appendix}

Here, we give a two-dimensional symmetric explicit representation of the conductivity matrix $\{\mathcal{L}_{\alpha\beta}\}$. This form is useful when the conditions of positive definiteness have to be calculated. Though the explicit writing is cumbersome, it is especially useful when an abstract notation is not sufficient, but explicit calculations must be done, or when a computer program for solving equations or carrying out numerical simulations must be implemented.  

\vspace{-1cm}

\subsection*{Representation of the conductivity matrix $\{\mathcal{L}_{\alpha\beta}\}$ in the case where the internal variable Q has odd parity}

Entropy production \eqref{eqn:36} of Subsection \ref{4.2}, can be  written in the symbolic matrix notation
\begin{equation}
\label{N}
X_\alpha \mathcal{L}_{\alpha\beta}X_{\beta}\geq 0, 
\end{equation}
where
\begin{equation}
\label{eqn:IXT}
\begin{split}
\{X_\alpha\}&=
\{
q_i \ ; \ q_{i,j} \ ; \ Q_{ij} \ ; \ Q_{ij,p} 
\}= \\
& =\{ q_1 \ ; \  q_2 \ ; \  q_3 \ ; \  q_{1,1} \ ; \  q_{1,2} \ ; \  q_{1,3} \ ; \  q_{2,1} \ ; \  q_{2,2} \ ; \  q_{2,3} \ ; \  q_{3,1} \ ; \  q_{3,2} \ ; \  q_{3,3} \ ; \\
& \qquad Q_{11,1} \ ; \  Q_{11,2} \ ; \  Q_{11,3} \ ; \  Q_{12,1} \ ; \  Q_{12,2} \ ; \  Q_{12,3} \ ; \  Q_{13,1} \ ; \  Q_{13,2} \ ; \  Q_{13,3} \ ; \\
& \qquad Q_{21,1} \ ; \  Q_{21,2} \ ; \  Q_{21,3} \ ; \  Q_{22,1} \ ; \  Q_{22,2} \ ; \  Q_{22,3} \ ; \  Q_{23,1} \ ; \  Q_{23,2} \ ; \  Q_{23,3} \ ; \\
& \qquad Q_{31,1} \ ; \  Q_{31,2} \ ; \  Q_{31,3} \ ; \  Q_{32,1} \ ; \  Q_{32,2} \ ; \  Q_{32,3} \ ; \  Q_{33,1} \ ; \  Q_{33,2} \ ; \  Q_{33,3} \}, \\[0.5em]
& \quad (\alpha=1,\ldots,48),
\end{split}
\end{equation}

\begin{equation}
\label{eqn:IX}
\{X_\beta\}=
\begin{Bmatrix}
q_k \\[0.5em]
q_{k,l} \\[0.5em]
Q_{kl} \\[0.5em]
Q_{lm,n} 
\end{Bmatrix},
\quad (\beta=1,\ldots,48).
\end{equation}

\noindent For $\mathcal{L}_{\alpha\beta}$ we introduce the following notation

\begin{equation}
\label{eqn:Imatrix}
\{\mathcal{L}_{\alpha\beta}\}=
\left(\begin{array}{@{}c|c|c|c@{}}
\overset{3\times 3}{\mathcal{L}^{(1)}_{ik}} & \overset{3\times 9}{0} & \overset{3\times 9}{0} & \overset{3\times 27}{\mathcal{L}^{(1,4)}_{ilmn}} \\[0.5em]
\hline \overset{9\times 3}{0} & \overset{}{\overset{9\times 9}{\mathcal{L}^{(2)}_{jikl}}} & \overset{9\times 9}{\mathcal{L}^{(2,3)}_{jikl}} & \overset{9\times 27}{0} \\[0.5em]
\hline \overset{9\times 3}{0} & \overset{}{\overset{9\times 9}{\mathcal{L}^{(3,2)}_{ijkl}}} & \overset{9\times 9}{\mathcal{L}^{(3)}_{ijkl}} & \overset{9\times 27}{0} \\[0.5em]
\hline \overset{}{\overset{27\times 3}{\mathcal{L}^{(4,1)}_{kpij}}} & \overset{27\times 9}{0} & \overset{27\times 9}{0} & \overset{27\times 27}{\mathcal{L}^{(4)}_{pijlmn}} 
\end{array}\right)
\quad (\alpha , \beta=1,\ldots,48), 
\end{equation}

\noindent in which $\overset{n\times m}{0}$ is the symbolic null matrix of dimension $n\times m$. This matrix is symmetric by virtue of Onsager relations \eqref{eqn:O-C2}$_2$ and \eqref{eqn:O-C3}$_2$.

In the following we write the sub-matrices that appear in \eqref{eqn:Imatrix}
\begin{equation}
\label{157}
\mathcal{L}^{(1)}_{ik}=
\begin{pmatrix}
L^{(1)} & 0 & 0 \\
0 & L^{(1)} & 0 \\
0 & 0 & L^{(1)}
\end{pmatrix},
\end{equation}

\begin{equation}
\qquad \mathcal{L}^{(1,4)}_{ilmn}=
\begin{pmatrix}
\mathcal{L}^{(1,4)}_{1111} & \mathcal{L}^{(1,4)}_{2111} & \mathcal{L}^{(1,4)}_{3111} \\[0.5em]
\mathcal{L}^{(1,4)}_{1112} & \mathcal{L}^{(1,4)}_{2112} & \mathcal{L}^{(1,4)}_{3112} \\[0.5em] 
\mathcal{L}^{(1,4)}_{1113} & \mathcal{L}^{(1,4)}_{2113} & \mathcal{L}^{(1,4)}_{3113} \\[0.5em] 
\mathcal{L}^{(1,4)}_{1121} & \mathcal{L}^{(1,4)}_{2121} & \mathcal{L}^{(1,4)}_{3121} \\[0.5em]
\mathcal{L}^{(1,4)}_{1122} & \mathcal{L}^{(1,4)}_{2122} & \mathcal{L}^{(1,4)}_{3122} \\[0.5em]
\mathcal{L}^{(1,4)}_{1123} & \mathcal{L}^{(1,4)}_{2123} & \mathcal{L}^{(1,4)}_{3123} \\[0.5em] 
\mathcal{L}^{(1,4)}_{1131} & \mathcal{L}^{(1,4)}_{2131} & \mathcal{L}^{(1,4)}_{3131} \\[0.5em] 
\mathcal{L}^{(1,4)}_{1132} & \mathcal{L}^{(1,4)}_{2132} & \mathcal{L}^{(1,4)}_{3132} \\[0.5em]
\mathcal{L}^{(1,4)}_{1133} & \mathcal{L}^{(1,4)}_{2133} & \mathcal{L}^{(1,4)}_{3133} \\[0.5em]
\mathcal{L}^{(1,4)}_{1211} & \mathcal{L}^{(1,4)}_{2211} & \mathcal{L}^{(1,4)}_{3211} \\[0.5em]
\mathcal{L}^{(1,4)}_{1212} & \mathcal{L}^{(1,4)}_{2212} & \mathcal{L}^{(1,4)}_{3212} \\[0.5em]
\mathcal{L}^{(1,4)}_{1213} & \mathcal{L}^{(1,4)}_{2213} & \mathcal{L}^{(1,4)}_{3213} \\[0.5em]
\mathcal{L}^{(1,4)}_{1221} & \mathcal{L}^{(1,4)}_{2221} & \mathcal{L}^{(1,4)}_{3221} \\[0.5em]
\mathcal{L}^{(1,4)}_{1222} & \mathcal{L}^{(1,4)}_{2222} & \mathcal{L}^{(1,4)}_{3222} \\[0.5em]
\mathcal{L}^{(1,4)}_{1223} & \mathcal{L}^{(1,4)}_{2223} & \mathcal{L}^{(1,4)}_{3223} \\[0.5em]
\mathcal{L}^{(1,4)}_{1231} & \mathcal{L}^{(1,4)}_{2231} & \mathcal{L}^{(1,4)}_{3231} \\[0.5em] 
\mathcal{L}^{(1,4)}_{1232} & \mathcal{L}^{(1,4)}_{2232} & \mathcal{L}^{(1,4)}_{3232} \\[0.5em] 
\mathcal{L}^{(1,4)}_{1233} & \mathcal{L}^{(1,4)}_{2233} & \mathcal{L}^{(1,4)}_{3233} \\[0.5em]
\mathcal{L}^{(1,4)}_{1311} & \mathcal{L}^{(1,4)}_{2311} & \mathcal{L}^{(1,4)}_{3311} \\[0.5em]
\mathcal{L}^{(1,4)}_{1312} & \mathcal{L}^{(1,4)}_{2312} & \mathcal{L}^{(1,4)}_{3312} \\[0.5em]
\mathcal{L}^{(1,4)}_{1313} & \mathcal{L}^{(1,4)}_{2313} & \mathcal{L}^{(1,4)}_{3313} \\[0.5em]
\mathcal{L}^{(1,4)}_{1321} & \mathcal{L}^{(1,4)}_{2321} & \mathcal{L}^{(1,4)}_{3321} \\[0.5em]
\mathcal{L}^{(1,4)}_{1322} & \mathcal{L}^{(1,4)}_{2322} & \mathcal{L}^{(1,4)}_{3322} \\[0.5em]
\mathcal{L}^{(1,4)}_{1323} & \mathcal{L}^{(1,4)}_{2323} & \mathcal{L}^{(1,4)}_{3323} \\[0.5em]
\mathcal{L}^{(1,4)}_{1331} & \mathcal{L}^{(1,4)}_{2331} & \mathcal{L}^{(1,4)}_{3331} \\[0.5em]
\mathcal{L}^{(1,4)}_{1332} & \mathcal{L}^{(1,4)}_{2332} & \mathcal{L}^{(1,4)}_{3332} \\[0.5em] 
\mathcal{L}^{(1,4)}_{1333} & \mathcal{L}^{(1,4)}_{2333} & \mathcal{L}^{(1,4)}_{3333} 
\end{pmatrix}
^T=
\begin{pmatrix}
L^{(1,4)} & 0 & 0 \\[0.3em]
0 & L^{(1,4)}_3 & 0 \\[0.3em]
0 & 0 & L^{(1,4)}_3 \\[0.3em]
0 & L^{(1,4)}_2 & 0 \\[0.3em]
L^{(1,4)}_1 & 0 & 0 \\[0.3em]
0 & 0 & 0 \\[0.3em]
0 & 0 & L^{(1,4)}_2 \\[0.3em]
0 & 0 & 0 \\[0.3em]
L^{(1,4)}_1 & 0 & 0 \\[0.3em]
0 & L^{(1,4)}_1 & 0 \\[0.3em]
L^{(1,4)}_2 & 0 & 0 \\[0.3em]
0 & 0 & 0 \\[0.3em]
L^{(1,4)}_3 & 0 & 0 \\[0.3em]
0 & L^{(1,4)} & 0 \\[0.3em]
0 & 0 & L^{(1,4)}_3 \\[0.3em]
0 & 0 & 0 \\[0.3em]
0 & 0 & L^{(1,4)}_2 \\[0.3em]
0 & L^{(1,4)}_1 & 0 \\[0.3em]
0 & 0 & L^{(1,4)}_1 \\[0.3em]
0 & 0 & 0 \\[0.3em]
L^{(1,4)}_2 & 0 & 0 \\[0.3em]
0 & 0 & 0 \\[0.3em]
0 & 0 & L^{(1,4)}_1 \\[0.3em]
0 & L^{(1,4)}_2 & 0 \\[0.3em]
L^{(1,4)}_3 & 0 & 0 \\[0.3em]
0 & L^{(1,4)}_3 & 0 \\[0.3em]
0 & 0 & L^{(1,4)}
\end{pmatrix}
^T,
\end{equation}
where $L^{(1,4)} \equiv L^{(1,4)}_1+L^{(1,4)}_2+L^{(1,4)}_3$.

\begin{equation}
\begin{split}
\mathcal{L}^{(2)}_{jikl}&=
\begin{pmatrix}
\mathcal{L}^{(2)}_{1111} & \mathcal{L}^{(2)}_{1112} & \mathcal{L}^{(2)}_{1113} & \mathcal{L}^{(2)}_{1121} & \mathcal{L}^{(2)}_{1122} & \mathcal{L}^{(2)}_{1123} & \mathcal{L}^{(2)}_{1131} & \mathcal{L}^{(2)}_{1132} & \mathcal{L}^{(2)}_{1133} \\[0.5em] 
\mathcal{L}^{(2)}_{2111} & \mathcal{L}^{(2)}_{2112} & \mathcal{L}^{(2)}_{2113} & \mathcal{L}^{(2)}_{2121} & \mathcal{L}^{(2)}_{2122} & \mathcal{L}^{(2)}_{2123} & \mathcal{L}^{(2)}_{2131} & \mathcal{L}^{(2)}_{2132} & \mathcal{L}^{(2)}_{2133} \\[0.5em]  
\mathcal{L}^{(2)}_{3111} & \mathcal{L}^{(2)}_{3112} & \mathcal{L}^{(2)}_{3113} & \mathcal{L}^{(2)}_{3121} & \mathcal{L}^{(2)}_{3122} & \mathcal{L}^{(2)}_{3123} & \mathcal{L}^{(2)}_{3131} & \mathcal{L}^{(2)}_{3132} & \mathcal{L}^{(2)}_{3133} \\[0.5em]  
\mathcal{L}^{(2)}_{1211} & \mathcal{L}^{(2)}_{1212} & \mathcal{L}^{(2)}_{1213} & \mathcal{L}^{(2)}_{1221} & \mathcal{L}^{(2)}_{1222} & \mathcal{L}^{(2)}_{1223} & \mathcal{L}^{(2)}_{1231} & \mathcal{L}^{(2)}_{1232} & \mathcal{L}^{(2)}_{1233} \\[0.5em]  
\mathcal{L}^{(2)}_{2211} & \mathcal{L}^{(2)}_{2212} & \mathcal{L}^{(2)}_{2213} & \mathcal{L}^{(2)}_{2221} & \mathcal{L}^{(2)}_{2222} & \mathcal{L}^{(2)}_{2223} & \mathcal{L}^{(2)}_{2231} & \mathcal{L}^{(2)}_{2232} & \mathcal{L}^{(2)}_{2233} \\[0.5em]  
\mathcal{L}^{(2)}_{3211} & \mathcal{L}^{(2)}_{3212} & \mathcal{L}^{(2)}_{3213} & \mathcal{L}^{(2)}_{3221} & \mathcal{L}^{(2)}_{3222} & \mathcal{L}^{(2)}_{3223} & \mathcal{L}^{(2)}_{3231} & \mathcal{L}^{(2)}_{3232} & \mathcal{L}^{(2)}_{3233} \\[0.5em]  
\mathcal{L}^{(2)}_{1311} & \mathcal{L}^{(2)}_{1312} & \mathcal{L}^{(2)}_{1313} & \mathcal{L}^{(2)}_{1321} & \mathcal{L}^{(2)}_{1322} & \mathcal{L}^{(2)}_{1323} & \mathcal{L}^{(2)}_{1331} & \mathcal{L}^{(2)}_{1332} & \mathcal{L}^{(2)}_{1333} \\[0.5em]  
\mathcal{L}^{(2)}_{2311} & \mathcal{L}^{(2)}_{2312} & \mathcal{L}^{(2)}_{2313} & \mathcal{L}^{(2)}_{2321} & \mathcal{L}^{(2)}_{2322} & \mathcal{L}^{(2)}_{2323} & \mathcal{L}^{(2)}_{2331} & \mathcal{L}^{(2)}_{2332} & \mathcal{L}^{(2)}_{2333} \\[0.5em]  
\mathcal{L}^{(2)}_{3311} & \mathcal{L}^{(2)}_{3312} & \mathcal{L}^{(2)}_{3313} & \mathcal{L}^{(2)}_{3321} & \mathcal{L}^{(2)}_{3322} & \mathcal{L}^{(2)}_{3323} & \mathcal{L}^{(2)}_{3331} & \mathcal{L}^{(2)}_{3332} & \mathcal{L}^{(2)}_{3333} 
\end{pmatrix}
= \\[1em]
& =
\begin{pmatrix}
L^{(2)} & 0 & 0 & 0 & L^{(2)}_1 & 0 & 0 & 0 & L^{(2)}_1 \\
0 & L^{(2)}_3 & 0 & L^{(2)}_2 & 0 & 0 & 0 & 0 & 0 \\
0 & 0 & L^{(2)}_3 & 0 & 0 & 0 & L^{(2)}_2 & 0 & 0 \\
0 & L^{(2)}_2 & 0 & L^{(2)}_3 & 0 & 0 & 0 & 0 & 0 \\
L^{(2)}_1 & 0 & 0 & 0 & L^{(2)} & 0 & 0 & 0 & L^{(2)}_1 \\
0 & 0 & 0 & 0 & 0 & L^{(2)}_3 & 0 & L^{(2)}_2 & 0 \\
0 & 0 & L^{(2)}_2 & 0 & 0 & 0 & L^{(2)}_3 & 0 & 0 \\
0 & 0 & 0 & 0 & 0 & L^{(2)}_2 & 0 & L^{(2)}_3 & 0 \\
L^{(2)}_1 & 0 & 0 & 0 & L^{(2)}_1 & 0 & 0 & 0 & L^{(2)}
\end{pmatrix},
\end{split}
\end{equation}
where $L^{(2)} \equiv L^{(2)}_1+L^{(2)}_2+L^{(2)}_3$.

\begin{equation}
\begin{split}
\mathcal{L}^{(2,3)}_{jikl}&=
\begin{pmatrix}
\mathcal{L}^{(2,3)}_{1111} & \mathcal{L}^{(2,3)}_{1112} & \mathcal{L}^{(2,3)}_{1113} & \mathcal{L}^{(2,3)}_{1121} & \mathcal{L}^{(2,3)}_{1122} & \mathcal{L}^{(2,3)}_{1123} & \mathcal{L}^{(2,3)}_{1131} & \mathcal{L}^{(2,3)}_{1132} & \mathcal{L}^{(2,3)}_{1133} \\[0.5em]  
\mathcal{L}^{(2,3)}_{2111} & \mathcal{L}^{(2,3)}_{2112} & \mathcal{L}^{(2,3)}_{2113} & \mathcal{L}^{(2,3)}_{2121} & \mathcal{L}^{(2,3)}_{2122} & \mathcal{L}^{(2,3)}_{2123} & \mathcal{L}^{(2,3)}_{2131} & \mathcal{L}^{(2,3)}_{2132} & \mathcal{L}^{(2,3)}_{2133} \\[0.5em]  
\mathcal{L}^{(2,3)}_{3111} & \mathcal{L}^{(2,3)}_{3112} & \mathcal{L}^{(2,3)}_{3113} & \mathcal{L}^{(2,3)}_{3121} & \mathcal{L}^{(2,3)}_{3122} & \mathcal{L}^{(2,3)}_{3123} & \mathcal{L}^{(2,3)}_{3131} & \mathcal{L}^{(2,3)}_{3132} & \mathcal{L}^{(2,3)}_{3133} \\[0.5em]  
\mathcal{L}^{(2,3)}_{1211} & \mathcal{L}^{(2,3)}_{1212} & \mathcal{L}^{(2,3)}_{1213} & \mathcal{L}^{(2,3)}_{1221} & \mathcal{L}^{(2,3)}_{1222} & \mathcal{L}^{(2,3)}_{1223} & \mathcal{L}^{(2,3)}_{1231} & \mathcal{L}^{(2,3)}_{1232} & \mathcal{L}^{(2,3)}_{1233} \\[0.5em]  
\mathcal{L}^{(2,3)}_{2211} & \mathcal{L}^{(2,3)}_{2212} & \mathcal{L}^{(2,3)}_{2213} & \mathcal{L}^{(2,3)}_{2221} & \mathcal{L}^{(2,3)}_{2222} & \mathcal{L}^{(2,3)}_{2223} & \mathcal{L}^{(2,3)}_{2231} & \mathcal{L}^{(2,3)}_{2232} & \mathcal{L}^{(2,3)}_{2233} \\[0.5em]  
\mathcal{L}^{(2,3)}_{3211} & \mathcal{L}^{(2,3)}_{3212} & \mathcal{L}^{(2,3)}_{3213} & \mathcal{L}^{(2,3)}_{3221} & \mathcal{L}^{(2,3)}_{3222} & \mathcal{L}^{(2,3)}_{3223} & \mathcal{L}^{(2,3)}_{3231} & \mathcal{L}^{(2,3)}_{3232} & \mathcal{L}^{(2,3)}_{3233} \\[0.5em]  
\mathcal{L}^{(2,3)}_{1311} & \mathcal{L}^{(2,3)}_{1312} & \mathcal{L}^{(2,3)}_{1313} & \mathcal{L}^{(2,3)}_{1321} & \mathcal{L}^{(2,3)}_{1322} & \mathcal{L}^{(2,3)}_{1323} & \mathcal{L}^{(2,3)}_{1331} & \mathcal{L}^{(2,3)}_{1332} & \mathcal{L}^{(2,3)}_{1333} \\[0.5em]  
\mathcal{L}^{(2,3)}_{2311} & \mathcal{L}^{(2,3)}_{2312} & \mathcal{L}^{(2,3)}_{2313} & \mathcal{L}^{(2,3)}_{2321} & \mathcal{L}^{(2,3)}_{2322} & \mathcal{L}^{(2,3)}_{2323} & \mathcal{L}^{(2,3)}_{2331} & \mathcal{L}^{(2,3)}_{2332} & \mathcal{L}^{(2,3)}_{2333} \\[0.5em]  
\mathcal{L}^{(2,3)}_{3311} & \mathcal{L}^{(2,3)}_{3312} & \mathcal{L}^{(2,3)}_{3313} & \mathcal{L}^{(2,3)}_{3321} & \mathcal{L}^{(2,3)}_{3322} & \mathcal{L}^{(2,3)}_{3323} & \mathcal{L}^{(2,3)}_{3331} & \mathcal{L}^{(2,3)}_{3332} & \mathcal{L}^{(2,3)}_{3333} 
\end{pmatrix}
= \\[1em]
& =
\begin{pmatrix}
L^{(2,3)} & 0 & 0 & 0 & L^{(2,3)}_1 & 0 & 0 & 0 & L^{(2,3)}_1 \\
0 & L^{(2,3)}_3 & 0 & L^{(2,3)}_2 & 0 & 0 & 0 & 0 & 0 \\
0 & 0 & L^{(2,3)}_3 & 0 & 0 & 0 & L^{(2,3)}_2 & 0 & 0 \\
0 & L^{(2,3)}_2 & 0 & L^{(2,3)}_3 & 0 & 0 & 0 & 0 & 0 \\
L^{(2,3)}_1 & 0 & 0 & 0 & L^{(2,3)} & 0 & 0 & 0 & L^{(2,3)}_1 \\
0 & 0 & 0 & 0 & 0 & L^{(2,3)}_3 & 0 & L^{(2,3)}_2 & 0 \\
0 & 0 & L^{(2,3)}_2 & 0 & 0 & 0 & L^{(2,3)}_3 & 0 & 0 \\
0 & 0 & 0 & 0 & 0 & L^{(2,3)}_2 & 0 & L^{(2,3)}_3 & 0 \\
L^{(2,3)}_1 & 0 & 0 & 0 & L^{(2,3)}_1 & 0 & 0 & 0 & L^{(2,3)}
\end{pmatrix},
\end{split}
\end{equation}
where $L^{(2,3)} \equiv L^{(2,3)}_1+L^{(2,3)}_2+L^{(2,3)}_3$.

\begin{equation}
\label{161}
\begin{split}
\mathcal{L}^{(3)}_{ijkl}&=
\begin{pmatrix}
\mathcal{L}^{(3)}_{1111} & \mathcal{L}^{(3)}_{1112} & \mathcal{L}^{(3)}_{1113} & \mathcal{L}^{(3)}_{1121} & \mathcal{L}^{(3)}_{1122} & \mathcal{L}^{(3)}_{1123} & \mathcal{L}^{(3)}_{1131} & \mathcal{L}^{(3)}_{1132} & \mathcal{L}^{(3)}_{1133} \\[0.5em]  
\mathcal{L}^{(3)}_{1211} & \mathcal{L}^{(3)}_{1212} & \mathcal{L}^{(3)}_{1213} & \mathcal{L}^{(3)}_{1221} & \mathcal{L}^{(3)}_{1222} & \mathcal{L}^{(3)}_{1223} & \mathcal{L}^{(3)}_{1231} & \mathcal{L}^{(3)}_{1232} & \mathcal{L}^{(3)}_{1233} \\[0.5em]  
\mathcal{L}^{(3)}_{1311} & \mathcal{L}^{(3)}_{1312} & \mathcal{L}^{(3)}_{1313} & \mathcal{L}^{(3)}_{1321} & \mathcal{L}^{(3)}_{1322} & \mathcal{L}^{(3)}_{1323} & \mathcal{L}^{(3)}_{1331} & \mathcal{L}^{(3)}_{1332} & \mathcal{L}^{(3)}_{1333} \\[0.5em]  
\mathcal{L}^{(3)}_{2111} & \mathcal{L}^{(3)}_{2112} & \mathcal{L}^{(3)}_{2113} & \mathcal{L}^{(3)}_{2121} & \mathcal{L}^{(3)}_{2122} & \mathcal{L}^{(3)}_{2123} & \mathcal{L}^{(3)}_{2131} & \mathcal{L}^{(3)}_{2132} & \mathcal{L}^{(3)}_{2133} \\[0.5em]  
\mathcal{L}^{(3)}_{2211} & \mathcal{L}^{(3)}_{2212} & \mathcal{L}^{(3)}_{2213} & \mathcal{L}^{(3)}_{2221} & \mathcal{L}^{(3)}_{2222} & \mathcal{L}^{(3)}_{2223} & \mathcal{L}^{(3)}_{2231} & \mathcal{L}^{(3)}_{2232} & \mathcal{L}^{(3)}_{2233} \\[0.5em]  
\mathcal{L}^{(3)}_{2311} & \mathcal{L}^{(3)}_{2312} & \mathcal{L}^{(3)}_{2313} & \mathcal{L}^{(3)}_{2321} & \mathcal{L}^{(3)}_{2322} & \mathcal{L}^{(3)}_{2323} & \mathcal{L}^{(3)}_{2331} & \mathcal{L}^{(3)}_{2332} & \mathcal{L}^{(3)}_{2333} \\[0.5em]  
\mathcal{L}^{(3)}_{3111} & \mathcal{L}^{(3)}_{3112} & \mathcal{L}^{(3)}_{3113} & \mathcal{L}^{(3)}_{3121} & \mathcal{L}^{(3)}_{3122} & \mathcal{L}^{(3)}_{3123} & \mathcal{L}^{(3)}_{3131} & \mathcal{L}^{(3)}_{3132} & \mathcal{L}^{(3)}_{3133} \\[0.5em]  
\mathcal{L}^{(3)}_{3211} & \mathcal{L}^{(3)}_{3212} & \mathcal{L}^{(3)}_{3213} & \mathcal{L}^{(3)}_{3221} & \mathcal{L}^{(3)}_{3222} & \mathcal{L}^{(3)}_{3223} & \mathcal{L}^{(3)}_{3231} & \mathcal{L}^{(3)}_{3232} & \mathcal{L}^{(3)}_{3233} \\[0.5em]  
\mathcal{L}^{(3)}_{3311} & \mathcal{L}^{(3)}_{3312} & \mathcal{L}^{(3)}_{3313} & \mathcal{L}^{(3)}_{3321} & \mathcal{L}^{(3)}_{3322} & \mathcal{L}^{(3)}_{3323} & \mathcal{L}^{(3)}_{3331} & \mathcal{L}^{(3)}_{3332} & \mathcal{L}^{(3)}_{3333} 
\end{pmatrix}
= \\[1em]
& =
\begin{pmatrix}
L^{(3)} & 0 & 0 & 0 & L^{(3)}_1 & 0 & 0 & 0 & L^{(3)}_1 \\
0 & L^{(3)}_2 & 0 & L^{(3)}_3 & 0 & 0 & 0 & 0 & 0 \\
0 & 0 & L^{(3)}_2 & 0 & 0 & 0 & L^{(3)}_3 & 0 & 0 \\
0 & L^{(3)}_3 & 0 & L^{(3)}_2 & 0 & 0 & 0 & 0 & 0 \\
L^{(3)}_1 & 0 & 0 & 0 & L^{(3)} & 0 & 0 & 0 & L^{(3)}_1 \\
0 & 0 & 0 & 0 & 0 & L^{(3)}_2 & 0 & L^{(3)}_3 & 0 \\
0 & 0 & L^{(3)}_3 & 0 & 0 & 0 & L^{(3)}_2 & 0 & 0 \\
0 & 0 & 0 & 0 & 0 & L^{(3)}_3 & 0 & L^{(3)}_2 & 0 \\
L^{(3)}_1 & 0 & 0 & 0 & L^{(3)}_1 & 0 & 0 & 0 & L^{(3)}
\end{pmatrix},
\end{split}
\end{equation}
where $L^{(3)} \equiv L^{(3)}_1+L^{(3)}_2+L^{(3)}_3$.

\begin{equation}
\label{162}
\begin{split}
\mathcal{L}^{(3,2)}_{ijkl}&=
\begin{pmatrix}
\mathcal{L}^{(2,3)}_{1111} & \mathcal{L}^{(2,3)}_{1112} & \mathcal{L}^{(2,3)}_{1113} & \mathcal{L}^{(2,3)}_{1121} & \mathcal{L}^{(2,3)}_{1122} & \mathcal{L}^{(2,3)}_{1123} & \mathcal{L}^{(2,3)}_{1131} & \mathcal{L}^{(2,3)}_{1132} & \mathcal{L}^{(2,3)}_{1133} \\[0.5em]  
\mathcal{L}^{(2,3)}_{1211} & \mathcal{L}^{(2,3)}_{1212} & \mathcal{L}^{(2,3)}_{1213} & \mathcal{L}^{(2,3)}_{1221} & \mathcal{L}^{(2,3)}_{1222} & \mathcal{L}^{(2,3)}_{1223} & \mathcal{L}^{(2,3)}_{1231} & \mathcal{L}^{(2,3)}_{1232} & \mathcal{L}^{(2,3)}_{1233} \\[0.5em]  
\mathcal{L}^{(2,3)}_{1311} & \mathcal{L}^{(2,3)}_{1312} & \mathcal{L}^{(2,3)}_{1313} & \mathcal{L}^{(2,3)}_{1321} & \mathcal{L}^{(2,3)}_{1322} & \mathcal{L}^{(2,3)}_{1323} & \mathcal{L}^{(2,3)}_{1331} & \mathcal{L}^{(2,3)}_{1332} & \mathcal{L}^{(2,3)}_{1333} \\[0.5em]  
\mathcal{L}^{(2,3)}_{2111} & \mathcal{L}^{(2,3)}_{2112} & \mathcal{L}^{(2,3)}_{2113} & \mathcal{L}^{(2,3)}_{2121} & \mathcal{L}^{(2,3)}_{2122} & \mathcal{L}^{(2,3)}_{2123} & \mathcal{L}^{(2,3)}_{2131} & \mathcal{L}^{(2,3)}_{2132} & \mathcal{L}^{(2,3)}_{2133} \\[0.5em]  
\mathcal{L}^{(2,3)}_{2211} & \mathcal{L}^{(2,3)}_{2212} & \mathcal{L}^{(2,3)}_{2213} & \mathcal{L}^{(2,3)}_{2221} & \mathcal{L}^{(2,3)}_{2222} & \mathcal{L}^{(2,3)}_{2223} & \mathcal{L}^{(2,3)}_{2231} & \mathcal{L}^{(2,3)}_{2232} & \mathcal{L}^{(2,3)}_{2233} \\[0.5em]  
\mathcal{L}^{(2,3)}_{2311} & \mathcal{L}^{(2,3)}_{2312} & \mathcal{L}^{(2,3)}_{2313} & \mathcal{L}^{(2,3)}_{2321} & \mathcal{L}^{(2,3)}_{2322} & \mathcal{L}^{(2,3)}_{2323} & \mathcal{L}^{(2,3)}_{2331} & \mathcal{L}^{(2,3)}_{2332} & \mathcal{L}^{(2,3)}_{2333} \\[0.5em]  
\mathcal{L}^{(2,3)}_{3111} & \mathcal{L}^{(2,3)}_{3112} & \mathcal{L}^{(2,3)}_{3113} & \mathcal{L}^{(2,3)}_{3121} & \mathcal{L}^{(2,3)}_{3122} & \mathcal{L}^{(2,3)}_{3123} & \mathcal{L}^{(2,3)}_{3131} & \mathcal{L}^{(2,3)}_{3132} & \mathcal{L}^{(2,3)}_{3133} \\[0.5em]  
\mathcal{L}^{(2,3)}_{3211} & \mathcal{L}^{(2,3)}_{3212} & \mathcal{L}^{(2,3)}_{3213} & \mathcal{L}^{(2,3)}_{3221} & \mathcal{L}^{(2,3)}_{3222} & \mathcal{L}^{(2,3)}_{3223} & \mathcal{L}^{(2,3)}_{3231} & \mathcal{L}^{(2,3)}_{3232} & \mathcal{L}^{(2,3)}_{3233} \\[0.5em]  
\mathcal{L}^{(2,3)}_{3311} & \mathcal{L}^{(2,3)}_{3312} & \mathcal{L}^{(2,3)}_{3313} & \mathcal{L}^{(2,3)}_{3321} & \mathcal{L}^{(2,3)}_{3322} & \mathcal{L}^{(2,3)}_{3323} & \mathcal{L}^{(2,3)}_{3331} & \mathcal{L}^{(2,3)}_{3332} & \mathcal{L}^{(2,3)}_{3333} 
\end{pmatrix}
= \\[1em]
& =
\begin{pmatrix}
L^{(2,3)} & 0 & 0 & 0 & L^{(2,3)}_1 & 0 & 0 & 0 & L^{(2,3)}_1 \\
0 & L^{(2,3)}_2 & 0 & L^{(2,3)}_3 & 0 & 0 & 0 & 0 & 0 \\
0 & 0 & L^{(2,3)}_2 & 0 & 0 & 0 & L^{(2,3)}_3 & 0 & 0 \\
0 & L^{(2,3)}_3 & 0 & L^{(2,3)}_2 & 0 & 0 & 0 & 0 & 0 \\
L^{(2,3)}_1 & 0 & 0 & 0 & L^{(2,3)} & 0 & 0 & 0 & L^{(2,3)}_1 \\
0 & 0 & 0 & 0 & 0 & L^{(2,3)}_2 & 0 & L^{(2,3)}_3 & 0 \\
0 & 0 & L^{(2,3)}_3 & 0 & 0 & 0 & L^{(2,3)}_2 & 0 & 0 \\
0 & 0 & 0 & 0 & 0 & L^{(2,3)}_3 & 0 & L^{(2,3)}_2 & 0 \\
L^{(2,3)}_1 & 0 & 0 & 0 & L^{(2,3)}_1 & 0 & 0 & 0 & L^{(2,3)}
\end{pmatrix},
\end{split}
\end{equation}
where we have used the Onsager relations \eqref{eqn:O-C3}$_2$.

\begin{equation}
\label{163}
\qquad \mathcal{L}^{(4,1)}_{kpij}=
\begin{pmatrix}
\mathcal{L}^{(1,4)}_{1111} & \mathcal{L}^{(1,4)}_{2111} & \mathcal{L}^{(1,4)}_{3111} \\[0.5em]
\mathcal{L}^{(1,4)}_{1211} & \mathcal{L}^{(1,4)}_{2211} & \mathcal{L}^{(1,4)}_{3211} \\[0.5em] 
\mathcal{L}^{(1,4)}_{1311} & \mathcal{L}^{(1,4)}_{2311} & \mathcal{L}^{(1,4)}_{3311} \\[0.5em] 
\mathcal{L}^{(1,4)}_{1112} & \mathcal{L}^{(1,4)}_{2112} & \mathcal{L}^{(1,4)}_{3112} \\[0.5em]
\mathcal{L}^{(1,4)}_{1212} & \mathcal{L}^{(1,4)}_{2212} & \mathcal{L}^{(1,4)}_{3212} \\[0.5em]
\mathcal{L}^{(1,4)}_{1312} & \mathcal{L}^{(1,4)}_{2312} & \mathcal{L}^{(1,4)}_{3312} \\[0.5em] 
\mathcal{L}^{(1,4)}_{1113} & \mathcal{L}^{(1,4)}_{2113} & \mathcal{L}^{(1,4)}_{3113} \\[0.5em] 
\mathcal{L}^{(1,4)}_{1213} & \mathcal{L}^{(1,4)}_{2213} & \mathcal{L}^{(1,4)}_{3213} \\[0.5em]
\mathcal{L}^{(1,4)}_{1313} & \mathcal{L}^{(1,4)}_{2313} & \mathcal{L}^{(1,4)}_{3313} \\[0.5em]
\mathcal{L}^{(1,4)}_{1121} & \mathcal{L}^{(1,4)}_{2121} & \mathcal{L}^{(1,4)}_{3121} \\[0.5em]
\mathcal{L}^{(1,4)}_{1221} & \mathcal{L}^{(1,4)}_{2221} & \mathcal{L}^{(1,4)}_{3221} \\[0.5em]
\mathcal{L}^{(1,4)}_{1321} & \mathcal{L}^{(1,4)}_{2321} & \mathcal{L}^{(1,4)}_{3321} \\[0.5em]
\mathcal{L}^{(1,4)}_{1122} & \mathcal{L}^{(1,4)}_{2122} & \mathcal{L}^{(1,4)}_{3122} \\[0.5em]
\mathcal{L}^{(1,4)}_{1222} & \mathcal{L}^{(1,4)}_{2222} & \mathcal{L}^{(1,4)}_{3222} \\[0.5em]
\mathcal{L}^{(1,4)}_{1322} & \mathcal{L}^{(1,4)}_{2322} & \mathcal{L}^{(1,4)}_{3322} \\[0.5em]
\mathcal{L}^{(1,4)}_{1123} & \mathcal{L}^{(1,4)}_{2123} & \mathcal{L}^{(1,4)}_{3123} \\[0.5em] 
\mathcal{L}^{(1,4)}_{1223} & \mathcal{L}^{(1,4)}_{2223} & \mathcal{L}^{(1,4)}_{3223} \\[0.5em] 
\mathcal{L}^{(1,4)}_{1323} & \mathcal{L}^{(1,4)}_{2323} & \mathcal{L}^{(1,4)}_{3323} \\[0.5em]
\mathcal{L}^{(1,4)}_{1131} & \mathcal{L}^{(1,4)}_{2131} & \mathcal{L}^{(1,4)}_{3131} \\[0.5em]
\mathcal{L}^{(1,4)}_{1231} & \mathcal{L}^{(1,4)}_{2231} & \mathcal{L}^{(1,4)}_{3231} \\[0.5em]
\mathcal{L}^{(1,4)}_{1331} & \mathcal{L}^{(1,4)}_{2331} & \mathcal{L}^{(1,4)}_{3331} \\[0.5em]
\mathcal{L}^{(1,4)}_{1132} & \mathcal{L}^{(1,4)}_{2132} & \mathcal{L}^{(1,4)}_{3132} \\[0.5em]
\mathcal{L}^{(1,4)}_{1232} & \mathcal{L}^{(1,4)}_{2232} & \mathcal{L}^{(1,4)}_{3232} \\[0.5em]
\mathcal{L}^{(1,4)}_{1332} & \mathcal{L}^{(1,4)}_{2332} & \mathcal{L}^{(1,4)}_{3332} \\[0.5em]
\mathcal{L}^{(1,4)}_{1133} & \mathcal{L}^{(1,4)}_{2133} & \mathcal{L}^{(1,4)}_{3133} \\[0.5em]
\mathcal{L}^{(1,4)}_{1233} & \mathcal{L}^{(1,4)}_{2233} & \mathcal{L}^{(1,4)}_{3233} \\[0.5em] 
\mathcal{L}^{(1,4)}_{1333} & \mathcal{L}^{(1,4)}_{2333} & \mathcal{L}^{(1,4)}_{3333} 
\end{pmatrix}
=
\begin{pmatrix}
L^{(1,4)} & 0 & 0 \\[0.3em]
0 & L^{(1,4)}_1 & 0 \\[0.3em]
0 & 0 & L^{(1,4)}_1 \\[0.3em]
0 & L^{(1,4)}_3 & 0 \\[0.3em]
L^{(1,4)}_2 & 0 & 0 \\[0.3em]
0 & 0 & 0 \\[0.3em]
0 & 0 & L^{(1,4)}_3 \\[0.3em]
0 & 0 & 0 \\[0.3em]
L^{(1,4)}_2 & 0 & 0 \\[0.3em]
0 & L^{(1,4)}_2 & 0 \\[0.3em]
L^{(1,4)}_3 & 0 & 0 \\[0.3em]
0 & 0 & 0 \\[0.3em]
L^{(1,4)}_1 & 0 & 0 \\[0.3em]
0 & L^{(1,4)} & 0 \\[0.3em]
0 & 0 & L^{(1,4)}_1 \\[0.3em]
0 & 0 & 0 \\[0.3em]
0 & 0 & L^{(1,4)}_3 \\[0.3em]
0 & L^{(1,4)}_2 & 0 \\[0.3em]
0 & 0 & L^{(1,4)}_2 \\[0.3em]
0 & 0 & 0 \\[0.3em]
L^{(1,4)}_3 & 0 & 0 \\[0.3em]
0 & 0 & 0 \\[0.3em]
0 & 0 & L^{(1,4)}_2 \\[0.3em]
0 & L^{(1,4)}_3 & 0 \\[0.3em]
L^{(1,4)}_1 & 0 & 0 \\[0.3em]
0 & L^{(1,4)}_1 & 0 \\[0.3em]
0 & 0 & L^{(1,4)}
\end{pmatrix},
\end{equation}
where $L^{(1,4)} \equiv L^{(1,4)}_1+L^{(1,4)}_2+L^{(1,4)}_3$ and we have used the Onsager relations \eqref{eqn:O-C2}$_2$.

\newgeometry{top=3cm,bottom=2.5cm,left=2.5cm,right=2.5cm,heightrounded}

\begin{landscape}


\begin{equation*}
\scalebox{0.54}{%
$\mathcal{L}^{(4)}_{pijlmn}=
\begin{pmatrix} 
\mathcal{L}^{(4)}_{111111} & \mathcal{L}^{(4)}_{111112} & \mathcal{L}^{(4)}_{111113} & \mathcal{L}^{(4)}_{111121} & \mathcal{L}^{(4)}_{111122} & \mathcal{L}^{(4)}_{111123} & \mathcal{L}^{(4)}_{111131} & \mathcal{L}^{(4)}_{111132} & \mathcal{L}^{(4)}_{111133} & \mathcal{L}^{(4)}_{111211} & \mathcal{L}^{(4)}_{111212} & \mathcal{L}^{(4)}_{111213} & \mathcal{L}^{(4)}_{111221} & \mathcal{L}^{(4)}_{111222} & \mathcal{L}^{(4)}_{111223} & \mathcal{L}^{(4)}_{111231} & \mathcal{L}^{(4)}_{111232} & \mathcal{L}^{(4)}_{111233} & \mathcal{L}^{(4)}_{111311} & \mathcal{L}^{(4)}_{111312} & \mathcal{L}^{(4)}_{111313} & \mathcal{L}^{(4)}_{111321} & \mathcal{L}^{(4)}_{111322} & \mathcal{L}^{(4)}_{111323} & \mathcal{L}^{(4)}_{111331} & \mathcal{L}^{(4)}_{111332} & \mathcal{L}^{(4)}_{111333}  \\[0.5em] 
\mathcal{L}^{(4)}_{211111} & \mathcal{L}^{(4)}_{211112} & \mathcal{L}^{(4)}_{211113} & \mathcal{L}^{(4)}_{211121} & \mathcal{L}^{(4)}_{211122} & \mathcal{L}^{(4)}_{211123} & \mathcal{L}^{(4)}_{211131} & \mathcal{L}^{(4)}_{211132} & \mathcal{L}^{(4)}_{211133} & \mathcal{L}^{(4)}_{211211} & \mathcal{L}^{(4)}_{211212} & \mathcal{L}^{(4)}_{211213} & \mathcal{L}^{(4)}_{211221} & \mathcal{L}^{(4)}_{211222} & \mathcal{L}^{(4)}_{211223} & \mathcal{L}^{(4)}_{211231} & \mathcal{L}^{(4)}_{211232} & \mathcal{L}^{(4)}_{211233} & \mathcal{L}^{(4)}_{211311} & \mathcal{L}^{(4)}_{211312} & \mathcal{L}^{(4)}_{211313} & \mathcal{L}^{(4)}_{211321} & \mathcal{L}^{(4)}_{211322} & \mathcal{L}^{(4)}_{211323} & \mathcal{L}^{(4)}_{211331} & \mathcal{L}^{(4)}_{211332} & \mathcal{L}^{(4)}_{211333}  \\[0.5em] 
\mathcal{L}^{(4)}_{311111} & \mathcal{L}^{(4)}_{311112} & \mathcal{L}^{(4)}_{311113} & \mathcal{L}^{(4)}_{311121} & \mathcal{L}^{(4)}_{311122} & \mathcal{L}^{(4)}_{311123} & \mathcal{L}^{(4)}_{311131} & \mathcal{L}^{(4)}_{311132} & \mathcal{L}^{(4)}_{311133} & \mathcal{L}^{(4)}_{311211} & \mathcal{L}^{(4)}_{311212} & \mathcal{L}^{(4)}_{311213} & \mathcal{L}^{(4)}_{311221} & \mathcal{L}^{(4)}_{311222} & \mathcal{L}^{(4)}_{311223} & \mathcal{L}^{(4)}_{311231} & \mathcal{L}^{(4)}_{311232} & \mathcal{L}^{(4)}_{311233} & \mathcal{L}^{(4)}_{311311} & \mathcal{L}^{(4)}_{311312} & \mathcal{L}^{(4)}_{311313} & \mathcal{L}^{(4)}_{311321} & \mathcal{L}^{(4)}_{311322} & \mathcal{L}^{(4)}_{311323} & \mathcal{L}^{(4)}_{311331} & \mathcal{L}^{(4)}_{311332} & \mathcal{L}^{(4)}_{311333}  \\[0.5em] 
\mathcal{L}^{(4)}_{112111} & \mathcal{L}^{(4)}_{112112} & \mathcal{L}^{(4)}_{112113} & \mathcal{L}^{(4)}_{112121} & \mathcal{L}^{(4)}_{112122} & \mathcal{L}^{(4)}_{112123} & \mathcal{L}^{(4)}_{112131} & \mathcal{L}^{(4)}_{112132} & \mathcal{L}^{(4)}_{112133} & \mathcal{L}^{(4)}_{112211} & \mathcal{L}^{(4)}_{112212} & \mathcal{L}^{(4)}_{112213} & \mathcal{L}^{(4)}_{112221} & \mathcal{L}^{(4)}_{112222} & \mathcal{L}^{(4)}_{112223} & \mathcal{L}^{(4)}_{112231} & \mathcal{L}^{(4)}_{112232} & \mathcal{L}^{(4)}_{112233} & \mathcal{L}^{(4)}_{112311} & \mathcal{L}^{(4)}_{112312} & \mathcal{L}^{(4)}_{112313} & \mathcal{L}^{(4)}_{112321} & \mathcal{L}^{(4)}_{112322} & \mathcal{L}^{(4)}_{112323} & \mathcal{L}^{(4)}_{112331} & \mathcal{L}^{(4)}_{112332} & \mathcal{L}^{(4)}_{112333}  \\[0.5em] 
\mathcal{L}^{(4)}_{212111} & \mathcal{L}^{(4)}_{212112} & \mathcal{L}^{(4)}_{212113} & \mathcal{L}^{(4)}_{212121} & \mathcal{L}^{(4)}_{212122} & \mathcal{L}^{(4)}_{212123} & \mathcal{L}^{(4)}_{212131} & \mathcal{L}^{(4)}_{212132} & \mathcal{L}^{(4)}_{212133} & \mathcal{L}^{(4)}_{212211} & \mathcal{L}^{(4)}_{212212} & \mathcal{L}^{(4)}_{212213} & \mathcal{L}^{(4)}_{212221} & \mathcal{L}^{(4)}_{212222} & \mathcal{L}^{(4)}_{212223} & \mathcal{L}^{(4)}_{212231} & \mathcal{L}^{(4)}_{212232} & \mathcal{L}^{(4)}_{212233} & \mathcal{L}^{(4)}_{212311} & \mathcal{L}^{(4)}_{212312} & \mathcal{L}^{(4)}_{212313} & \mathcal{L}^{(4)}_{212321} & \mathcal{L}^{(4)}_{212322} & \mathcal{L}^{(4)}_{212323} & \mathcal{L}^{(4)}_{212331} & \mathcal{L}^{(4)}_{212332} & \mathcal{L}^{(4)}_{212333}  \\[0.5em] 
\mathcal{L}^{(4)}_{312111} & \mathcal{L}^{(4)}_{312112} & \mathcal{L}^{(4)}_{312113} & \mathcal{L}^{(4)}_{312121} & \mathcal{L}^{(4)}_{312122} & \mathcal{L}^{(4)}_{312123} & \mathcal{L}^{(4)}_{312131} & \mathcal{L}^{(4)}_{312132} & \mathcal{L}^{(4)}_{312133} & \mathcal{L}^{(4)}_{312211} & \mathcal{L}^{(4)}_{312212} & \mathcal{L}^{(4)}_{312213} & \mathcal{L}^{(4)}_{312221} & \mathcal{L}^{(4)}_{312222} & \mathcal{L}^{(4)}_{312223} & \mathcal{L}^{(4)}_{312231} & \mathcal{L}^{(4)}_{312232} & \mathcal{L}^{(4)}_{312233} & \mathcal{L}^{(4)}_{312311} & \mathcal{L}^{(4)}_{312312} & \mathcal{L}^{(4)}_{312313} & \mathcal{L}^{(4)}_{312321} & \mathcal{L}^{(4)}_{312322} & \mathcal{L}^{(4)}_{312323} & \mathcal{L}^{(4)}_{312331} & \mathcal{L}^{(4)}_{312332} & \mathcal{L}^{(4)}_{312333}  \\[0.5em] 
\mathcal{L}^{(4)}_{113111} & \mathcal{L}^{(4)}_{113112} & \mathcal{L}^{(4)}_{113113} & \mathcal{L}^{(4)}_{113121} & \mathcal{L}^{(4)}_{113122} & \mathcal{L}^{(4)}_{113123} & \mathcal{L}^{(4)}_{113131} & \mathcal{L}^{(4)}_{113132} & \mathcal{L}^{(4)}_{113133} & \mathcal{L}^{(4)}_{113211} & \mathcal{L}^{(4)}_{113212} & \mathcal{L}^{(4)}_{113213} & \mathcal{L}^{(4)}_{113221} & \mathcal{L}^{(4)}_{113222} & \mathcal{L}^{(4)}_{113223} & \mathcal{L}^{(4)}_{113231} & \mathcal{L}^{(4)}_{113232} & \mathcal{L}^{(4)}_{113233} & \mathcal{L}^{(4)}_{113311} & \mathcal{L}^{(4)}_{113312} & \mathcal{L}^{(4)}_{113313} & \mathcal{L}^{(4)}_{113321} & \mathcal{L}^{(4)}_{113322} & \mathcal{L}^{(4)}_{113323} & \mathcal{L}^{(4)}_{113331} & \mathcal{L}^{(4)}_{113332} & \mathcal{L}^{(4)}_{113333}  \\[0.5em] 
\mathcal{L}^{(4)}_{213111} & \mathcal{L}^{(4)}_{213112} & \mathcal{L}^{(4)}_{213113} & \mathcal{L}^{(4)}_{213121} & \mathcal{L}^{(4)}_{213122} & \mathcal{L}^{(4)}_{213123} & \mathcal{L}^{(4)}_{213131} & \mathcal{L}^{(4)}_{213132} & \mathcal{L}^{(4)}_{213133} & \mathcal{L}^{(4)}_{213211} & \mathcal{L}^{(4)}_{213212} & \mathcal{L}^{(4)}_{213213} & \mathcal{L}^{(4)}_{213221} & \mathcal{L}^{(4)}_{213222} & \mathcal{L}^{(4)}_{213223} & \mathcal{L}^{(4)}_{213231} & \mathcal{L}^{(4)}_{213232} & \mathcal{L}^{(4)}_{213233} & \mathcal{L}^{(4)}_{213311} & \mathcal{L}^{(4)}_{213312} & \mathcal{L}^{(4)}_{213313} & \mathcal{L}^{(4)}_{213321} & \mathcal{L}^{(4)}_{213322} & \mathcal{L}^{(4)}_{213323} & \mathcal{L}^{(4)}_{213331} & \mathcal{L}^{(4)}_{213332} & \mathcal{L}^{(4)}_{213333}  \\[0.5em] 
\mathcal{L}^{(4)}_{313111} & \mathcal{L}^{(4)}_{313112} & \mathcal{L}^{(4)}_{313113} & \mathcal{L}^{(4)}_{313121} & \mathcal{L}^{(4)}_{313122} & \mathcal{L}^{(4)}_{313123} & \mathcal{L}^{(4)}_{313131} & \mathcal{L}^{(4)}_{313132} & \mathcal{L}^{(4)}_{313133} & \mathcal{L}^{(4)}_{313211} & \mathcal{L}^{(4)}_{313212} & \mathcal{L}^{(4)}_{313213} & \mathcal{L}^{(4)}_{313221} & \mathcal{L}^{(4)}_{313222} & \mathcal{L}^{(4)}_{313223} & \mathcal{L}^{(4)}_{313231} & \mathcal{L}^{(4)}_{313232} & \mathcal{L}^{(4)}_{313233} & \mathcal{L}^{(4)}_{313311} & \mathcal{L}^{(4)}_{313312} & \mathcal{L}^{(4)}_{313313} & \mathcal{L}^{(4)}_{313321} & \mathcal{L}^{(4)}_{313322} & \mathcal{L}^{(4)}_{313323} & \mathcal{L}^{(4)}_{313331} & \mathcal{L}^{(4)}_{313332} & \mathcal{L}^{(4)}_{313333}  \\[0.5em] 
\mathcal{L}^{(4)}_{121111} & \mathcal{L}^{(4)}_{121112} & \mathcal{L}^{(4)}_{121113} & \mathcal{L}^{(4)}_{121121} & \mathcal{L}^{(4)}_{121122} & \mathcal{L}^{(4)}_{121123} & \mathcal{L}^{(4)}_{121131} & \mathcal{L}^{(4)}_{121132} & \mathcal{L}^{(4)}_{121133} & \mathcal{L}^{(4)}_{121211} & \mathcal{L}^{(4)}_{121212} & \mathcal{L}^{(4)}_{121213} & \mathcal{L}^{(4)}_{121221} & \mathcal{L}^{(4)}_{121222} & \mathcal{L}^{(4)}_{121223} & \mathcal{L}^{(4)}_{121231} & \mathcal{L}^{(4)}_{121232} & \mathcal{L}^{(4)}_{121233} & \mathcal{L}^{(4)}_{121311} & \mathcal{L}^{(4)}_{121312} & \mathcal{L}^{(4)}_{121313} & \mathcal{L}^{(4)}_{121321} & \mathcal{L}^{(4)}_{121322} & \mathcal{L}^{(4)}_{121323} & \mathcal{L}^{(4)}_{121331} & \mathcal{L}^{(4)}_{121332} & \mathcal{L}^{(4)}_{121333}  \\[0.5em] 
\mathcal{L}^{(4)}_{221111} & \mathcal{L}^{(4)}_{221112} & \mathcal{L}^{(4)}_{221113} & \mathcal{L}^{(4)}_{221121} & \mathcal{L}^{(4)}_{221122} & \mathcal{L}^{(4)}_{221123} & \mathcal{L}^{(4)}_{221131} & \mathcal{L}^{(4)}_{221132} & \mathcal{L}^{(4)}_{221133} & \mathcal{L}^{(4)}_{221211} & \mathcal{L}^{(4)}_{221212} & \mathcal{L}^{(4)}_{221213} & \mathcal{L}^{(4)}_{221221} & \mathcal{L}^{(4)}_{221222} & \mathcal{L}^{(4)}_{221223} & \mathcal{L}^{(4)}_{221231} & \mathcal{L}^{(4)}_{221232} & \mathcal{L}^{(4)}_{221233} & \mathcal{L}^{(4)}_{221311} & \mathcal{L}^{(4)}_{221312} & \mathcal{L}^{(4)}_{221313} & \mathcal{L}^{(4)}_{221321} & \mathcal{L}^{(4)}_{221322} & \mathcal{L}^{(4)}_{221323} & \mathcal{L}^{(4)}_{221331} & \mathcal{L}^{(4)}_{221332} & \mathcal{L}^{(4)}_{221333}  \\[0.5em] 
\mathcal{L}^{(4)}_{321111} & \mathcal{L}^{(4)}_{321112} & \mathcal{L}^{(4)}_{321113} & \mathcal{L}^{(4)}_{321121} & \mathcal{L}^{(4)}_{321122} & \mathcal{L}^{(4)}_{321123} & \mathcal{L}^{(4)}_{321131} & \mathcal{L}^{(4)}_{321132} & \mathcal{L}^{(4)}_{321133} & \mathcal{L}^{(4)}_{321211} & \mathcal{L}^{(4)}_{321212} & \mathcal{L}^{(4)}_{321213} & \mathcal{L}^{(4)}_{321221} & \mathcal{L}^{(4)}_{321222} & \mathcal{L}^{(4)}_{321223} & \mathcal{L}^{(4)}_{321231} & \mathcal{L}^{(4)}_{321232} & \mathcal{L}^{(4)}_{321233} & \mathcal{L}^{(4)}_{321311} & \mathcal{L}^{(4)}_{321312} & \mathcal{L}^{(4)}_{321313} & \mathcal{L}^{(4)}_{321321} & \mathcal{L}^{(4)}_{321322} & \mathcal{L}^{(4)}_{321323} & \mathcal{L}^{(4)}_{321331} & \mathcal{L}^{(4)}_{321332} & \mathcal{L}^{(4)}_{321333}  \\[0.5em] 
\mathcal{L}^{(4)}_{122111} & \mathcal{L}^{(4)}_{122112} & \mathcal{L}^{(4)}_{122113} & \mathcal{L}^{(4)}_{122121} & \mathcal{L}^{(4)}_{122122} & \mathcal{L}^{(4)}_{122123} & \mathcal{L}^{(4)}_{122131} & \mathcal{L}^{(4)}_{122132} & \mathcal{L}^{(4)}_{122133} & \mathcal{L}^{(4)}_{122211} & \mathcal{L}^{(4)}_{122212} & \mathcal{L}^{(4)}_{122213} & \mathcal{L}^{(4)}_{122221} & \mathcal{L}^{(4)}_{122222} & \mathcal{L}^{(4)}_{122223} & \mathcal{L}^{(4)}_{122231} & \mathcal{L}^{(4)}_{122232} & \mathcal{L}^{(4)}_{122233} & \mathcal{L}^{(4)}_{122311} & \mathcal{L}^{(4)}_{122312} & \mathcal{L}^{(4)}_{122313} & \mathcal{L}^{(4)}_{122321} & \mathcal{L}^{(4)}_{122322} & \mathcal{L}^{(4)}_{122323} & \mathcal{L}^{(4)}_{122331} & \mathcal{L}^{(4)}_{122332} & \mathcal{L}^{(4)}_{122333}  \\[0.5em] 
\mathcal{L}^{(4)}_{222111} & \mathcal{L}^{(4)}_{222112} & \mathcal{L}^{(4)}_{222113} & \mathcal{L}^{(4)}_{222121} & \mathcal{L}^{(4)}_{222122} & \mathcal{L}^{(4)}_{222123} & \mathcal{L}^{(4)}_{222131} & \mathcal{L}^{(4)}_{222132} & \mathcal{L}^{(4)}_{222133} & \mathcal{L}^{(4)}_{222211} & \mathcal{L}^{(4)}_{222212} & \mathcal{L}^{(4)}_{222213} & \mathcal{L}^{(4)}_{222221} & \mathcal{L}^{(4)}_{222222} & \mathcal{L}^{(4)}_{222223} & \mathcal{L}^{(4)}_{222231} & \mathcal{L}^{(4)}_{222232} & \mathcal{L}^{(4)}_{222233} & \mathcal{L}^{(4)}_{222311} & \mathcal{L}^{(4)}_{222312} & \mathcal{L}^{(4)}_{222313} & \mathcal{L}^{(4)}_{222321} & \mathcal{L}^{(4)}_{222322} & \mathcal{L}^{(4)}_{222323} & \mathcal{L}^{(4)}_{222331} & \mathcal{L}^{(4)}_{222332} & \mathcal{L}^{(4)}_{222333}   \\[0.5em] 
\mathcal{L}^{(4)}_{322111} & \mathcal{L}^{(4)}_{322112} & \mathcal{L}^{(4)}_{322113} & \mathcal{L}^{(4)}_{322121} & \mathcal{L}^{(4)}_{322122} & \mathcal{L}^{(4)}_{322123} & \mathcal{L}^{(4)}_{322131} & \mathcal{L}^{(4)}_{322132} & \mathcal{L}^{(4)}_{322133} & \mathcal{L}^{(4)}_{322211} & \mathcal{L}^{(4)}_{322212} & \mathcal{L}^{(4)}_{322213} & \mathcal{L}^{(4)}_{322221} & \mathcal{L}^{(4)}_{322222} & \mathcal{L}^{(4)}_{322223} & \mathcal{L}^{(4)}_{322231} & \mathcal{L}^{(4)}_{322232} & \mathcal{L}^{(4)}_{322233} & \mathcal{L}^{(4)}_{322311} & \mathcal{L}^{(4)}_{322312} & \mathcal{L}^{(4)}_{322313} & \mathcal{L}^{(4)}_{322321} & \mathcal{L}^{(4)}_{322322} & \mathcal{L}^{(4)}_{322323} & \mathcal{L}^{(4)}_{322331} & \mathcal{L}^{(4)}_{322332} & \mathcal{L}^{(4)}_{322333}   \\[0.5em] 
\mathcal{L}^{(4)}_{123111} & \mathcal{L}^{(4)}_{123112} & \mathcal{L}^{(4)}_{123113} & \mathcal{L}^{(4)}_{123121} & \mathcal{L}^{(4)}_{123122} & \mathcal{L}^{(4)}_{123123} & \mathcal{L}^{(4)}_{123131} & \mathcal{L}^{(4)}_{123132} & \mathcal{L}^{(4)}_{123133} & \mathcal{L}^{(4)}_{123211} & \mathcal{L}^{(4)}_{123212} & \mathcal{L}^{(4)}_{123213} & \mathcal{L}^{(4)}_{123221} & \mathcal{L}^{(4)}_{123222} & \mathcal{L}^{(4)}_{123223} & \mathcal{L}^{(4)}_{123231} & \mathcal{L}^{(4)}_{123232} & \mathcal{L}^{(4)}_{123233} & \mathcal{L}^{(4)}_{123311} & \mathcal{L}^{(4)}_{123312} & \mathcal{L}^{(4)}_{123313} & \mathcal{L}^{(4)}_{123321} & \mathcal{L}^{(4)}_{123322} & \mathcal{L}^{(4)}_{123323} & \mathcal{L}^{(4)}_{123331} & \mathcal{L}^{(4)}_{123332} & \mathcal{L}^{(4)}_{123333}   \\[0.5em] 
\mathcal{L}^{(4)}_{223111} & \mathcal{L}^{(4)}_{223112} & \mathcal{L}^{(4)}_{223113} & \mathcal{L}^{(4)}_{223121} & \mathcal{L}^{(4)}_{223122} & \mathcal{L}^{(4)}_{223123} & \mathcal{L}^{(4)}_{223131} & \mathcal{L}^{(4)}_{223132} & \mathcal{L}^{(4)}_{223133} & \mathcal{L}^{(4)}_{223211} & \mathcal{L}^{(4)}_{223212} & \mathcal{L}^{(4)}_{223213} & \mathcal{L}^{(4)}_{223221} & \mathcal{L}^{(4)}_{223222} & \mathcal{L}^{(4)}_{223223} & \mathcal{L}^{(4)}_{223231} & \mathcal{L}^{(4)}_{223232} & \mathcal{L}^{(4)}_{223233} & \mathcal{L}^{(4)}_{223311} & \mathcal{L}^{(4)}_{223312} & \mathcal{L}^{(4)}_{223313} & \mathcal{L}^{(4)}_{223321} & \mathcal{L}^{(4)}_{223322} & \mathcal{L}^{(4)}_{223323} & \mathcal{L}^{(4)}_{223331} & \mathcal{L}^{(4)}_{223332} & \mathcal{L}^{(4)}_{223333}  \\[0.5em] 
\mathcal{L}^{(4)}_{323111} & \mathcal{L}^{(4)}_{323112} & \mathcal{L}^{(4)}_{323113} & \mathcal{L}^{(4)}_{323121} & \mathcal{L}^{(4)}_{323122} & \mathcal{L}^{(4)}_{323123} & \mathcal{L}^{(4)}_{323131} & \mathcal{L}^{(4)}_{323132} & \mathcal{L}^{(4)}_{323133} & \mathcal{L}^{(4)}_{323211} & \mathcal{L}^{(4)}_{323212} & \mathcal{L}^{(4)}_{323213} & \mathcal{L}^{(4)}_{323221} & \mathcal{L}^{(4)}_{323222} & \mathcal{L}^{(4)}_{323223} & \mathcal{L}^{(4)}_{323231} & \mathcal{L}^{(4)}_{323232} & \mathcal{L}^{(4)}_{323233} & \mathcal{L}^{(4)}_{323311} & \mathcal{L}^{(4)}_{323312} & \mathcal{L}^{(4)}_{323313} & \mathcal{L}^{(4)}_{323321} & \mathcal{L}^{(4)}_{323322} & \mathcal{L}^{(4)}_{323323} & \mathcal{L}^{(4)}_{323331} & \mathcal{L}^{(4)}_{323332} & \mathcal{L}^{(4)}_{323333}  \\[0.5em] 
\mathcal{L}^{(4)}_{131111} & \mathcal{L}^{(4)}_{131112} & \mathcal{L}^{(4)}_{131113} & \mathcal{L}^{(4)}_{131121} & \mathcal{L}^{(4)}_{131122} & \mathcal{L}^{(4)}_{131123} & \mathcal{L}^{(4)}_{131131} & \mathcal{L}^{(4)}_{131132} & \mathcal{L}^{(4)}_{131133} & \mathcal{L}^{(4)}_{131211} & \mathcal{L}^{(4)}_{131212} & \mathcal{L}^{(4)}_{131213} & \mathcal{L}^{(4)}_{131221} & \mathcal{L}^{(4)}_{131222} & \mathcal{L}^{(4)}_{131223} & \mathcal{L}^{(4)}_{131231} & \mathcal{L}^{(4)}_{131232} & \mathcal{L}^{(4)}_{131233} & \mathcal{L}^{(4)}_{131311} & \mathcal{L}^{(4)}_{131312} & \mathcal{L}^{(4)}_{131313} & \mathcal{L}^{(4)}_{131321} & \mathcal{L}^{(4)}_{131322} & \mathcal{L}^{(4)}_{131323} & \mathcal{L}^{(4)}_{131331} & \mathcal{L}^{(4)}_{131332} & \mathcal{L}^{(4)}_{131333}  \\[0.5em] 
\mathcal{L}^{(4)}_{231111} & \mathcal{L}^{(4)}_{231112} & \mathcal{L}^{(4)}_{231113} & \mathcal{L}^{(4)}_{231121} & \mathcal{L}^{(4)}_{231122} & \mathcal{L}^{(4)}_{231123} & \mathcal{L}^{(4)}_{231131} & \mathcal{L}^{(4)}_{231132} & \mathcal{L}^{(4)}_{231133} & \mathcal{L}^{(4)}_{231211} & \mathcal{L}^{(4)}_{231212} & \mathcal{L}^{(4)}_{231213} & \mathcal{L}^{(4)}_{231221} & \mathcal{L}^{(4)}_{231222} & \mathcal{L}^{(4)}_{231223} & \mathcal{L}^{(4)}_{231231} & \mathcal{L}^{(4)}_{231232} & \mathcal{L}^{(4)}_{231233} & \mathcal{L}^{(4)}_{231311} & \mathcal{L}^{(4)}_{231312} & \mathcal{L}^{(4)}_{231313} & \mathcal{L}^{(4)}_{231321} & \mathcal{L}^{(4)}_{231322} & \mathcal{L}^{(4)}_{231323} & \mathcal{L}^{(4)}_{231331} & \mathcal{L}^{(4)}_{231332} & \mathcal{L}^{(4)}_{231333}  \\[0.5em] 
\mathcal{L}^{(4)}_{331111} & \mathcal{L}^{(4)}_{331112} & \mathcal{L}^{(4)}_{331113} & \mathcal{L}^{(4)}_{331121} & \mathcal{L}^{(4)}_{331122} & \mathcal{L}^{(4)}_{331123} & \mathcal{L}^{(4)}_{331131} & \mathcal{L}^{(4)}_{331132} & \mathcal{L}^{(4)}_{331133} & \mathcal{L}^{(4)}_{331211} & \mathcal{L}^{(4)}_{331212} & \mathcal{L}^{(4)}_{331213} & \mathcal{L}^{(4)}_{331221} & \mathcal{L}^{(4)}_{331222} & \mathcal{L}^{(4)}_{331223} & \mathcal{L}^{(4)}_{331231} & \mathcal{L}^{(4)}_{331232} & \mathcal{L}^{(4)}_{331233} & \mathcal{L}^{(4)}_{331311} & \mathcal{L}^{(4)}_{331312} & \mathcal{L}^{(4)}_{331313} & \mathcal{L}^{(4)}_{331321} & \mathcal{L}^{(4)}_{331322} & \mathcal{L}^{(4)}_{331323} & \mathcal{L}^{(4)}_{331331} & \mathcal{L}^{(4)}_{331332} & \mathcal{L}^{(4)}_{331333}  \\[0.5em] 
\mathcal{L}^{(4)}_{132111} & \mathcal{L}^{(4)}_{132112} & \mathcal{L}^{(4)}_{132113} & \mathcal{L}^{(4)}_{132121} & \mathcal{L}^{(4)}_{132122} & \mathcal{L}^{(4)}_{132123} & \mathcal{L}^{(4)}_{132131} & \mathcal{L}^{(4)}_{132132} & \mathcal{L}^{(4)}_{132133} & \mathcal{L}^{(4)}_{132211} & \mathcal{L}^{(4)}_{132212} & \mathcal{L}^{(4)}_{132213} & \mathcal{L}^{(4)}_{132221} & \mathcal{L}^{(4)}_{132222} & \mathcal{L}^{(4)}_{132223} & \mathcal{L}^{(4)}_{132231} & \mathcal{L}^{(4)}_{132232} & \mathcal{L}^{(4)}_{132233} & \mathcal{L}^{(4)}_{132311} & \mathcal{L}^{(4)}_{132312} & \mathcal{L}^{(4)}_{132313} & \mathcal{L}^{(4)}_{132321} & \mathcal{L}^{(4)}_{132322} & \mathcal{L}^{(4)}_{132323} & \mathcal{L}^{(4)}_{132331} & \mathcal{L}^{(4)}_{132332} & \mathcal{L}^{(4)}_{132333}  \\[0.5em] 
\mathcal{L}^{(4)}_{232111} & \mathcal{L}^{(4)}_{232112} & \mathcal{L}^{(4)}_{232113} & \mathcal{L}^{(4)}_{232121} & \mathcal{L}^{(4)}_{232122} & \mathcal{L}^{(4)}_{232123} & \mathcal{L}^{(4)}_{232131} & \mathcal{L}^{(4)}_{232132} & \mathcal{L}^{(4)}_{232133} & \mathcal{L}^{(4)}_{232211} & \mathcal{L}^{(4)}_{232212} & \mathcal{L}^{(4)}_{232213} & \mathcal{L}^{(4)}_{232221} & \mathcal{L}^{(4)}_{232222} & \mathcal{L}^{(4)}_{232223} & \mathcal{L}^{(4)}_{232231} & \mathcal{L}^{(4)}_{232232} & \mathcal{L}^{(4)}_{232233} & \mathcal{L}^{(4)}_{232311} & \mathcal{L}^{(4)}_{232312} & \mathcal{L}^{(4)}_{232313} & \mathcal{L}^{(4)}_{232321} & \mathcal{L}^{(4)}_{232322} & \mathcal{L}^{(4)}_{232323} & \mathcal{L}^{(4)}_{232331} & \mathcal{L}^{(4)}_{232332} & \mathcal{L}^{(4)}_{232333}  \\[0.5em] 
\mathcal{L}^{(4)}_{332111} & \mathcal{L}^{(4)}_{332112} & \mathcal{L}^{(4)}_{332113} & \mathcal{L}^{(4)}_{332121} & \mathcal{L}^{(4)}_{332122} & \mathcal{L}^{(4)}_{332123} & \mathcal{L}^{(4)}_{332131} & \mathcal{L}^{(4)}_{332132} & \mathcal{L}^{(4)}_{332133} & \mathcal{L}^{(4)}_{332211} & \mathcal{L}^{(4)}_{332212} & \mathcal{L}^{(4)}_{332213} & \mathcal{L}^{(4)}_{332221} & \mathcal{L}^{(4)}_{332222} & \mathcal{L}^{(4)}_{332223} & \mathcal{L}^{(4)}_{332231} & \mathcal{L}^{(4)}_{332232} & \mathcal{L}^{(4)}_{332233} & \mathcal{L}^{(4)}_{332311} & \mathcal{L}^{(4)}_{332312} & \mathcal{L}^{(4)}_{332313} & \mathcal{L}^{(4)}_{332321} & \mathcal{L}^{(4)}_{332322} & \mathcal{L}^{(4)}_{332323} & \mathcal{L}^{(4)}_{332331} & \mathcal{L}^{(4)}_{332332} & \mathcal{L}^{(4)}_{332333}  \\[0.5em] 
\mathcal{L}^{(4)}_{133111} & \mathcal{L}^{(4)}_{133112} & \mathcal{L}^{(4)}_{133113} & \mathcal{L}^{(4)}_{133121} & \mathcal{L}^{(4)}_{133122} & \mathcal{L}^{(4)}_{133123} & \mathcal{L}^{(4)}_{133131} & \mathcal{L}^{(4)}_{133132} & \mathcal{L}^{(4)}_{133133} & \mathcal{L}^{(4)}_{133211} & \mathcal{L}^{(4)}_{133212} & \mathcal{L}^{(4)}_{133213} & \mathcal{L}^{(4)}_{133221} & \mathcal{L}^{(4)}_{133222} & \mathcal{L}^{(4)}_{133223} & \mathcal{L}^{(4)}_{133231} & \mathcal{L}^{(4)}_{133232} & \mathcal{L}^{(4)}_{133233} & \mathcal{L}^{(4)}_{133311} & \mathcal{L}^{(4)}_{133312} & \mathcal{L}^{(4)}_{133313} & \mathcal{L}^{(4)}_{133321} & \mathcal{L}^{(4)}_{133322} & \mathcal{L}^{(4)}_{133323} & \mathcal{L}^{(4)}_{133331} & \mathcal{L}^{(4)}_{133332} & \mathcal{L}^{(4)}_{133333}   \\[0.5em] 
\mathcal{L}^{(4)}_{233111} & \mathcal{L}^{(4)}_{233112} & \mathcal{L}^{(4)}_{233113} & \mathcal{L}^{(4)}_{233121} & \mathcal{L}^{(4)}_{233122} & \mathcal{L}^{(4)}_{233123} & \mathcal{L}^{(4)}_{233131} & \mathcal{L}^{(4)}_{233132} & \mathcal{L}^{(4)}_{233133} & \mathcal{L}^{(4)}_{233211} & \mathcal{L}^{(4)}_{233212} & \mathcal{L}^{(4)}_{233213} & \mathcal{L}^{(4)}_{233221} & \mathcal{L}^{(4)}_{233222} & \mathcal{L}^{(4)}_{233223} & \mathcal{L}^{(4)}_{233231} & \mathcal{L}^{(4)}_{233232} & \mathcal{L}^{(4)}_{233233} & \mathcal{L}^{(4)}_{233311} & \mathcal{L}^{(4)}_{233312} & \mathcal{L}^{(4)}_{233313} & \mathcal{L}^{(4)}_{233321} & \mathcal{L}^{(4)}_{233322} & \mathcal{L}^{(4)}_{233323} & \mathcal{L}^{(4)}_{233331} & \mathcal{L}^{(4)}_{233332} & \mathcal{L}^{(4)}_{233333}  \\[0.5em] 
\mathcal{L}^{(4)}_{333111} & \mathcal{L}^{(4)}_{333112} & \mathcal{L}^{(4)}_{333113} & \mathcal{L}^{(4)}_{333121} & \mathcal{L}^{(4)}_{333122} & \mathcal{L}^{(4)}_{333123} & \mathcal{L}^{(4)}_{333131} & \mathcal{L}^{(4)}_{333132} & \mathcal{L}^{(4)}_{333133} & \mathcal{L}^{(4)}_{333211} & \mathcal{L}^{(4)}_{333212} & \mathcal{L}^{(4)}_{333213} & \mathcal{L}^{(4)}_{333221} & \mathcal{L}^{(4)}_{333222} & \mathcal{L}^{(4)}_{333223} & \mathcal{L}^{(4)}_{333231} & \mathcal{L}^{(4)}_{333232} & \mathcal{L}^{(4)}_{333233} & \mathcal{L}^{(4)}_{333311} & \mathcal{L}^{(4)}_{333312} & \mathcal{L}^{(4)}_{333313} & \mathcal{L}^{(4)}_{333321} & \mathcal{L}^{(4)}_{333322} & \mathcal{L}^{(4)}_{333323} & \mathcal{L}^{(4)}_{333331} & \mathcal{L}^{(4)}_{333332} & \mathcal{L}^{(4)}_{333333} 
\end{pmatrix}
=$}
\end{equation*}


\newpage


\begin{equation}
\label{eqn:BigMatrix}
\scalebox{0.73}{%
$=
\begin{pmatrix} 
C^{(4)} & 0 & 0 & 0 & \Lambda_1 & 0 & 0 & 0 & \Lambda_1 & 0 & \Lambda_2 & 0 & \Lambda_3 & 0 & 0 & 0 & 0 & 0 & 0 & 0 & \Lambda_2 & 0 & 0 & 0 & \Lambda_3 & 0 & 0  \\[0.5em] 
0 & \Lambda_4 & 0 & \Lambda_5 & 0 & 0 & 0 & 0 & 0 & \Lambda_6 & 0 & 0 & 0 & \Lambda_1 & 0 & 0 & 0 & C^{(4)}_6 & 0 & 0 & 0 & 0 & 0 & C^{(4)}_4 & 0 & C^{(4)}_1 & 0  \\[0.5em] 
0 & 0 & \Lambda_4 & 0 & 0 & 0 & \Lambda_5 & 0 & 0 & 0 & 0 & 0 & 0 & 0 & C^{(4)}_1 & 0 & C^{(4)}_4 & 0 & \Lambda_6 & 0 & 0 & 0 & C^{(4)}_6 & 0 & 0 & 0 & \Lambda_1  \\[0.5em] 
0 & \Lambda_7 & 0 & \Lambda_8 & 0 & 0 & 0 & 0 & 0 & \Lambda_4 & 0 & 0 & 0 & \Lambda_3 & 0 & 0 & 0 & C^{(4)}_1 & 0 & 0 & 0 & 0 & 0 & C^{(4)}_2 & 0 & C^{(4)}_3 & 0  \\[0.5em] 
\Lambda_2 & 0 & 0 & 0 & \Lambda_5 & 0 & 0 & 0 & C^{(4)}_4 & 0 & \Lambda_9 & 0 & \Lambda_8 & 0 & 0 & 0 & 0 & 0 & 0 & 0 & C^{(4)}_5 & 0 & 0 & 0 & C^{(4)}_2 & 0 & 0  \\[0.5em] 
0 & 0 & 0 & 0 & 0 & C^{(4)}_{10} & 0 & C^{(4)}_9 & 0 & 0 & 0 & C^{(4)}_{11} & 0 & 0 & 0 & C^{(4)}_{10} & 0 & 0 & 0 & C^{(4)}_7 & 0 & C^{(4)}_8 & 0 & 0 & 0 & 0 & 0  \\[0.5em] 
0 & 0 & \Lambda_7 & 0 & 0 & 0 & \Lambda_8 & 0 & 0 & 0 & 0 & 0 & 0 & 0 & C^{(4)}_3 & 0 & C^{(4)}_2 & 0 & \Lambda_4 & 0 & 0 & 0 & C^{(4)}_1 & 0 & 0 & 0 & \Lambda_3  \\[0.5em] 
0 & 0 & 0 & 0 & 0 & C^{(4)}_9 & 0 & C^{(4)}_{10} & 0 & 0 & 0 & C^{(4)}_7 & 0 & 0 & 0 & C^{(4)}_8 & 0 & 0 & 0 & C^{(4)}_{11} & 0 & C^{(4)}_{10} & 0 & 0 & 0 & 0 & 0  \\[0.5em] 
\Lambda_2 & 0 & 0 & 0 & C^{(4)}_4 & 0 & 0 & 0 & \Lambda_5 & 0 & C^{(4)}_5 & 0 & C^{(4)}_2 & 0 & 0 & 0 & 0 & 0 & 0 & 0 & \Lambda_9 & 0 & 0 & 0 & \Lambda_8 & 0 & 0  \\[0.5em] 
0 & \Lambda_8 & 0 & \Lambda_9 & 0 & 0 & 0 & 0 & 0 & \Lambda_5 & 0 & 0 & 0 & \Lambda_2 & 0 & 0 & 0 & C^{(4)}_4 & 0 & 0 & 0 & 0 & 0 & C^{(4)}_5 & 0 & C^{(4)}_2 & 0  \\[0.5em] 
\Lambda_3 & 0 & 0 & 0 & \Lambda_4 & 0 & 0 & 0 & C^{(4)}_1 & 0 & \Lambda_8 & 0 & \Lambda_7 & 0 & 0 & 0 & 0 & 0 & 0 & 0 & C^{(4)}_2 & 0 & 0 & 0 & C^{(4)}_3 & 0 & 0  \\[0.5em] 
0 & 0 & 0 & 0 & 0 & C^{(4)}_{11} & 0 & C^{(4)}_{10} & 0 & 0 & 0 & C^{(4)}_{10} & 0 & 0 & 0 & C^{(4)}_9 & 0 & 0 & 0 & C^{(4)}_8 & 0 & C^{(4)}_7 & 0 & 0 & 0 & 0 & 0  \\[0.5em] 
\Lambda_1 & 0 & 0 & 0 & \Lambda_6 & 0 & 0 & 0 & C^{(4)}_6 & 0 & \Lambda_5 & 0 & \Lambda_4 & 0 & 0 & 0 & 0 & 0 & 0 & 0 & C^{(4)}_4 & 0 & 0 & 0 & C^{(4)}_1 & 0 & 0  \\[0.5em] 
0 & \Lambda_3 & 0 & \Lambda_2 & 0 & 0 & 0 & 0 & 0 & \Lambda_1 & 0 & 0 & 0 & C^{(4)} & 0 & 0 & 0 & \Lambda_1 & 0 & 0 & 0 & 0 & 0 & \Lambda_2 & 0 & \Lambda_3 & 0   \\[0.5em] 
0 & 0 & C^{(4)}_1 & 0 & 0 & 0 & C^{(4)}_4 & 0 & 0 & 0 & 0 & 0 & 0 & 0 & \Lambda_4 & 0 & \Lambda_5 & 0 & C^{(4)}_6 & 0 & 0 & 0 & \Lambda_6 & 0 & 0 & 0 & \Lambda_1   \\[0.5em] 
0 & 0 & 0 & 0 & 0 & C^{(4)}_7 & 0 & C^{(4)}_8 & 0 & 0 & 0 & C^{(4)}_9 & 0 & 0 & 0 & C^{(4)}_{10} & 0 & 0 & 0 & C^{(4)}_{10} & 0 & C^{(4)}_{11} & 0 & 0 & 0 & 0 & 0   \\[0.5em] 
0 & 0 & C^{(4)}_3 & 0 & 0 & 0 & C^{(4)}_2 & 0 & 0 & 0 & 0 & 0 & 0 & 0 & \Lambda_7 & 0 & \Lambda_8 & 0 & C^{(4)}_1 & 0 & 0 & 0 & \Lambda_4 & 0 & 0 & 0 & \Lambda_3  \\[0.5em] 
0 & C^{(4)}_2 & 0 & C^{(4)}_5 & 0 & 0 & 0 & 0 & 0 & C^{(4)}_4 & 0 & 0 & 0 & \Lambda_2 & 0 & 0 & 0 & \Lambda_5 & 0 & 0 & 0 & 0 & 0 & \Lambda_9 & 0 & \Lambda_8 & 0  \\[0.5em] 
0 & 0 & \Lambda_8 & 0 & 0 & 0 & \Lambda_9 & 0 & 0 & 0 & 0 & 0 & 0 & 0 & C^{(4)}_2 & 0 & C^{(4)}_5 & 0 & \Lambda_5 & 0 & 0 & 0 & C^{(4)}_4 & 0 & 0 & 0 & \Lambda_2  \\[0.5em] 
0 & 0 & 0 & 0 & 0 & C^{(4)}_{10} & 0 & C^{(4)}_{11} & 0 & 0 & 0 & C^{(4)}_8 & 0 & 0 & 0 & C^{(4)}_7 & 0 & 0 & 0 & C^{(4)}_{10} & 0 & C^{(4)}_9 & 0 & 0 & 0 & 0 & 0  \\[0.5em] 
\Lambda_3 & 0 & 0 & 0 & C^{(4)}_1 & 0 & 0 & 0 & \Lambda_4 & 0 & C^{(4)}_2 & 0 & C^{(4)}_3 & 0 & 0 & 0 & 0 & 0 & 0 & 0 & \Lambda_8 & 0 & 0 & 0 & \Lambda_7 & 0 & 0  \\[0.5em] 
0 & 0 & 0 & 0 & 0 & C^{(4)}_8 & 0 & C^{(4)}_7 & 0 & 0 & 0 & C^{(4)}_{10} & 0 & 0 & 0 & C^{(4)}_{11} & 0 & 0 & 0 & C^{(4)}_9 & 0 & C^{(4)}_{10} & 0 & 0 & 0 & 0 & 0  \\[0.5em] 
0 & 0 & C^{(4)}_2 & 0 & 0 & 0 & C^{(4)}_5 & 0 & 0 & 0 & 0 & 0 & 0 & 0 & \Lambda_8 & 0 & \Lambda_9 & 0 & C^{(4)}_4 & 0 & 0 & 0 & \Lambda_5 & 0 & 0 & 0 & \Lambda_2  \\[0.5em] 
0 & C^{(4)}_3 & 0 & C^{(4)}_2 & 0 & 0 & 0 & 0 & 0 & C^{(4)}_1 & 0 & 0 & 0 & \Lambda_3 & 0 & 0 & 0 & \Lambda_4 & 0 & 0 & 0 & 0 & 0 & \Lambda_8 & 0 & \Lambda_7 & 0  \\[0.5em] 
\Lambda_1 & 0 & 0 & 0 & C^{(4)}_6 & 0 & 0 & 0 & \Lambda_6 & 0 & C^{(4)}_4 & 0 & C^{(4)}_1 & 0 & 0 & 0 & 0 & 0 & 0 & 0 & \Lambda_5 & 0 & 0 & 0 & \Lambda_4 & 0 & 0   \\[0.5em] 
0 & C^{(4)}_1 & 0 & C^{(4)}_4 & 0 & 0 & 0 & 0 & 0 & C^{(4)}_6 & 0 & 0 & 0 & \Lambda_1 & 0 & 0 & 0 & \Lambda_6 & 0 & 0 & 0 & 0 & 0 & \Lambda_5 & 0 & \Lambda_4 & 0  \\[0.5em] 
0 & 0 & \Lambda_3 & 0 & 0 & 0 & \Lambda_2 & 0 & 0 & 0 & 0 & 0 & 0 & 0 & \Lambda_3 & 0 & \Lambda_2 & 0 & \Lambda_1 & 0 & 0 & 0 & \Lambda_1 & 0 & 0 & 0 & C^{(4)}
\end{pmatrix}$},
\end{equation}


\end{landscape}

\newpage
\restoregeometry 

where 
\begin{gather*}
C^{(4)}=2C^{(4)}_1+2C^{(4)}_2+C^{(4)}_3+2C^{(4)}_4+C^{(4)}_5+C^{(4)}_6+C^{(4)}_7+C^{(4)}_8+ C^{(4)}_9+2C^{(4)}_{10}+C^{(4)}_{11}, \\
\Lambda_1=C^{(4)}_1+C^{(4)}_4+C^{(4)}_6, \quad \Lambda_2=C^{(4)}_2+C^{(4)}_4+C^{(4)}_5, \quad \Lambda_3=C^{(4)}_1+C^{(4)}_2+C^{(4)}_3, \\
\Lambda_4=C^{(4)}_1+C^{(4)}_{10}+C^{(4)}_{11}, \quad \Lambda_5=C^{(4)}_4+C^{(4)}_9+C^{(4)}_{10}, \quad \Lambda_6=C^{(4)}_6+C^{(4)}_7+C^{(4)}_8, \\
\Lambda_7=C^{(4)}_3+C^{(4)}_7+C^{(4)}_9, \quad \Lambda_8=C^{(4)}_2+C^{(4)}_8+C^{(4)}_{10}, \quad \Lambda_9=C^{(4)}_5+C^{(4)}_7+C^{(4)}_{11}.
\end{gather*}

\subsection*{Representation of the conductivity matrix $\{\mathcal{L}_{\alpha\beta}\}$ in the case where  Q has even parity}

When the internal variable \textbf{Q} has even parity the expressions \eqref{N}-\eqref{eqn:Imatrix} and the results  \eqref{157}-\eqref{161} and \eqref{eqn:BigMatrix} of this Appendix remain unchanged. The phenomenological tensors $\mathcal{L}^{(3,2)}_{ijkl}$ and $\mathcal{L}^{(4,1)}_{kpij}$ are defined in the same way, but in the last terms of \eqref{162} and \eqref{163} before their matrix representation a minus sign appears because of in the calculations we take into account Onsager symmetry relations \eqref{eqn:O-C2}$_2$ \eqref{eqn:O-C3}$_2$. Thus, we have obtained that in the case where \textbf{Q} has odd parity the matrix $\{\mathcal{L}_{\alpha\beta}\}$ is symmetric, but in the case where Q has even parity  the matrix $\{\mathcal{L}_{\alpha\beta}\}$ is not symmetric because of Onsager relations \eqref{eqn:O-C2}$_2$ and \eqref{eqn:O-C3}$_2$.

\bibliographystyle{unsrt}

\end{document}